\documentclass[a4paper,12pt]{article}
\usepackage{Stylefile}

\onehalfspace

\title
{ %
\vspace*{5.0cm} \LARGE{\bf Modeling the Flow of a \BauMan\ Fluid in Porous Media} \vspace*{4.0cm} \\
}

\author{Taha Sochi\footnote{University College London - Department of Physics \& Astronomy - Gower Street - London. Email:
t.sochi@ucl.ac.uk.} \vspace*{3.0cm}}

\date{2009}

\begin{document}

\maketitle %
\pagenumbering{arabic}


\newpage
\phantomsection \addcontentsline{toc}{section}{Abstract}


\noindent
{\LARGE \bf \vspace{1.0cm} \\ Abstract} \vspace{0.5cm}\\
\noindent %
In this article, the extensional flow and viscosity and the \convdiv\ geometry were examined as the
basis of the peculiar \vc\ behavior in porous media. The modified \BauMan\ model, which
successfully describes \shThin, elasticity and \thixotropic\ time-dependency, was used for modeling
the flow of \vc\ materials which also show \thixotropic\ attributes. An algorithm, originally
proposed by Philippe Tardy, that employs this model to simulate \steadys\ \timedep\ flow was
implemented in a \nNEW\ flow simulation code using pore-scale modeling and the initial results were
analyzed. The findings are encouraging for further future development.


\newpage
\phantomsection \addcontentsline{toc}{section}{Contents} %
\tableofcontents

\newpage
\phantomsection \addcontentsline{toc}{section}{List of Figures} %
\listoffigures

\phantomsection \addcontentsline{toc}{section}{List of Tables} %
\listoftables

{\setlength{\parskip}{6pt plus 2pt minus 1pt}

\pagestyle{headings} %
\addtolength{\headheight}{+1.6pt}
\lhead[{Chapter \thechapter \thepage}]%
      {{\bfseries\rightmark}}
\rhead[{\bfseries\leftmark}]%
     {{\bfseries\thepage}} 
\headsep = 1.0cm               

\newpage
\section{Introduction} \label{Introduction}
\NNEW\ fluids are commonly divided into three broad groups \cite{skellandbook, ChhabraR1999}:
\begin{enumerate}

\item \Timeind: those fluids for which the strain rate at a given point is solely
dependent upon the instantaneous stress at that point.

\item \Vc: those fluids that show partial elastic recovery upon the removal of a
deforming stress. 

\item \Timedep: those fluids for which the strain rate is a function of both the
magnitude and the duration of stress and possibly of the time lapse between consecutive
applications of stress.

\end{enumerate}

A large number of models have been proposed in the literature to model all types of \nNEW\ fluids
under various flow conditions. Most these models are basically empirical in nature and arising from
curve-fitting exercises \cite{barnesbookHW1993}. In this article, we investigate a \nNEW\
rheological model (namely \BauMan\ model) that can be used to describe \vc\ behavior with
\thixotropic\ attributes in the context of pore-scale modeling of \nNEW\ flow through porous media.

\newpage
\section{\Vc\ Fluids} \label{}
\Vc\ substances exhibit a dual nature of behavior by showing signs of both viscous fluids and
elastic solids. In its most simple form, \vy\ can be modeled by combining \Newton's law for viscous
fluids (stress $\propto$ rate of strain) with \Hook's law for elastic solids (stress $\propto$
strain), as given by the original \Maxwell\ model and extended by the Convected \Maxwell\ models
for the nonlinear \vc\ fluids. Although this idealization predicts several
basic \vc\ phenomena, it does so only qualitatively \cite{larsonbook1988}.%

Polymeric fluids often show strong \vc\ effects, which can include \shThin, extension thickening,
\vc\ normal stresses, and \timedep\ rheology phenomena. The equations describing the flow of \vc\
fluids consist of the basic laws of continuum mechanics and the rheological equation of state, or
constitutive equation, describing a particular fluid and relates the \vc\ stress to the deformation
history. The quest is to derive a model that is as simple as possible, involving the minimum number
of variables and parameters, and yet having the capability to predict the \vc\ behavior in complex
flows \cite{larsonbook1999}.

No theory is yet available that can adequately describe all of the observed \vc\ phenomena in a
variety of flows. However, many differential and integral \vc\ constitutive models have been
proposed in the literature. What is common to all these is the presence of at least one
characteristic time parameter to account for the fluid memory, that is the stress at the present
time depends upon the strain or rate-of-strain for all past times, but with an exponentially fading
memory \cite{Hulsen1996-2, Keunings2004, owensbook2002, Denn1990, deiberthesis}.

The behavior of \vc\ fluids is drastically different from that of \NEW\ and inelastic \nNEW\
fluids. This includes the presence of normal stresses in shear flows, sensitivity to deformation
type, and memory effects such as stress relaxation and \timedep\ viscosity. These features underlie
the observed peculiar \vc\ phenomena such as rod-climbing (\Weissenberg\ effect), die swell and
open-channel siphon \cite{Boger1987, larsonbook1988}. Most \vc\ fluids exhibit \shThin\ and an
elongational viscosity that is both strain and extensional strain rate dependent, in contrast to
\NEW\ fluids where the elongational viscosity is constant \cite{Boger1987}.

The behavior of \vc\ fluids at any time is dependent on their recent deformation history, that is
they possess a fading memory of their past. Indeed a material that has no memory cannot be elastic,
since it has no way of remembering its original shape. Consequently, an ideal \vc\ fluid should
behave as an elastic solid in sufficiently rapid deformations and as a \NEW\ liquid in sufficiently
slow deformations. The justification is that the larger the strain rate, the more strain is imposed
on the sample within the memory span of the fluid \cite{Boger1987, birdbook, larsonbook1988}.

Many materials are \vc\ but at different time scales that may not be reached. Dependent on the time
scale of the flow, \vc\ materials mainly show viscous or elastic behavior. The particular response
of a sample in a given experiment depends on the time scale of the experiment in relation to a
natural time of the material. Thus, if the experiment is relatively slow, the sample will appear to
be viscous rather than elastic, whereas, if the experiment is relatively fast, it will appear to be
elastic rather than viscous. At intermediate time scales mixed \vc\ response is observed. Therefore
the concept of a natural time of a material is important in characterizing the material as viscous
or elastic. The ratio between the material time scale and the time scale of the flow is indicated
by a non-dimensional number: the \Deborah\ or the \Weissenberg\ number \cite{barnesbookHW1993,
wapperomthesis}.

A common feature of \vc\ fluids is stress relaxation after a sudden shearing displacement where
stress overshoots to a maximum then starts decreasing exponentially and eventually settles to a
steady state value. This phenomenon also takes place on cessation of steady shear flow where stress
decays over a finite measurable length of time. This reveals that \vc\ fluids are able to store and
release energy in contrast to inelastic fluids which react instantaneously to the imposed
deformation \cite{birdbook, deiberthesis, larsonbook1988}.

A defining characteristic of \vc\ materials associated with stress relaxation is the relaxation
time which may be defined as the time required for the shear stress in a simple shear flow to
return to zero under constant strain condition. Hence for a \Hookean\ elastic solid the relaxation
time is infinite, while for a \NEW\ fluid the relaxation of the stress is immediate and the
relaxation time is zero. Relaxation times which are infinite or zero are never realized in reality
as they correspond to the mathematical idealization of \Hookean\ elastic solids and \NEW\ liquids.
In practice, stress relaxation after the imposition of constant strain condition takes place over
some finite non-zero time interval \cite{owensbook2002}.

The complexity of \vy\ is phenomenal and the subject is notorious for being extremely difficult and
challenging. The constitutive equations for \vc\ fluids are much too complex to be treated in a
general manner. Further complications arise from the confusion created by the presence of other
phenomena such as wall effects and polymer-wall interactions, and these appear to be system
specific. Therefore, it is doubtful that a general fluid model capable of predicting all the flow
responses of \vc\ system with enough mathematical simplicity or tractability can emerge in the
foreseeable future \cite{Wissler1971, deiberthesis, ChhabraCM2001}. Understandably, despite the
huge amount of literature composed in the last few decades on this subject, almost all these
studies have investigated very simple cases in which substantial simplifications have been made
using basic \vc\ models.

In the last few decades, a general consensus has emerged that in the flow of \vc\ fluids through
porous media elastic effects should arise, though their precise nature is unknown or controversial.
In porous media, \vc\ effects can be important in certain cases. When they are, the actual pressure
gradient will exceed the purely viscous gradient beyond a critical flow rate, as observed by
several investigators. The normal stresses of high molecular polymer solutions can explain in part
the high flow resistances encountered during \vc\ flow through porous media. It is argued that the
very high normal stress differences and \TR\ ratios (defined as extensional to shear viscosity)
associated with polymeric fluids will produce increasing values of apparent viscosity when flow
channels in the porous medium are of rapidly changing cross section.

Important aspects of \nNEW\ flow in general and \vc\ flow in particular through porous media are
still presenting serious challenge for modeling and quantification. There are intrinsic
difficulties of characterizing \nNEW\ effects in the flow of polymer solutions and the complexities
of the local geometry of the porous medium which give rise to a complex and pore-space-dependent
flow field in which shear and extension coexist in various proportions that cannot be quantified.
Flows through porous media cannot be classified as pure shear flows as the \convdiv\ passages
impose a predominantly extensional flow fields especially at high flow rates. Moreover, the
extension viscosity of many \nNEW\ fluids increases dramatically with the extension rate. As a
consequence, the relationship between the pressure drop and flow rate very often do not follow the
observed \NEW\ and inelastic \nNEW\ trend. Further complication arises from the fact that for
complex fluids the stress depends not only on whether the flow is a shearing, extensional, or mixed
type, but also on the whole history of the velocity gradient \cite{MarshallM1967, DaubenM1967,
PearsonT2002, PlogK, MendesN2002, larsonbook1999}.

\section{Important Aspects for \Vc\ Flow in Porous Media} \label{ImportantAspects}
Strong experimental evidence indicates that the flow of \vc\ fluids through packed beds can exhibit
rapid increases in the pressure drop, or an increase in the apparent viscosity, above that expected
for a comparable purely viscous fluid. This increase has been attributed to the extensional nature
of the flow field in the pores caused by the successive expansions and contractions that a fluid
element experiences as it traverses the pore space. Even though the flow field at pore level is not
an ideal extensional flow due to the presence of shear and rotation, the increase in flow
resistance is referred to as an extension thickening effect \cite{ThienK1987, PlogK, DeiberS1981,
PilitsisB1989}.

There are two major interrelated aspects that have strong impact on the flow through porous media.
These are extensional flow and \convdiv\ geometry.

\subsection{Extensional Flow}\label{ExtensionalFlow}
One complexity in the flow in general and through porous media in particular usually arises from
the coexistence of shear and extensional components; sometimes with the added complication of
inertia. Pure shear or elongational flow is very much the exception in practical situations,
especially in the flow through porous media. By far the most common situation is for mixed flow to
occur where deformation rates have components parallel and perpendicular to the principal flow
direction. In such flows, the elongational components may be associated with the \convdiv\ flow
paths \cite{barnesbookHW1993, sorbiebook}.

A general consensus has emerged recently that the flow through packed beds has a substantial
extensional component and typical polymer solutions exhibit strain hardening in extension, which is
mainly responsible for the reported dramatic increases in pressure drop. Thus in principle the
shear viscosity alone is inadequate to explain the observed excessive pressure gradients. It is
therefore essential to know the relative importance of elastic and viscous effects or equivalently
the relationship between normal and shear stresses for different shear rates \cite{carreaubook,
ChhabraCM2001, deiberthesis}.

Elongational flow is fundamentally different from shear, the material property characterizing the
flow is not the viscosity, but the elongational viscosity. The behavior of the extensional
viscosity function is very often qualitatively different from that of the shear viscosity function.
For example, highly elastic polymer solutions that possess a viscosity which decreases
monotonically in shear often exhibit an extensional viscosity that increases dramatically with
strain rate. Thus, while the shear viscosity is \shThin\, the extensional viscosity is extension
thickening \cite{larsonbook1999, barnesbookHW1993, wapperomthesis}.

\subsection{\ConvDiv\ Geometry} \label{ConvergingDiverging}
An important aspect that characterizes the flow in porous media and makes it distinct from bulk is
the presence of \convdiv\ flow paths. This geometric factor significantly affects the flow and
accentuate elastic responses. Several arguments have been presented in the literature to explain
the effect of \convdiv\ geometry on the flow behavior. Despite this diversity, there is a general
consensus that in porous media the \convdiv\ nature of the flow paths brings out both the
extensional and shear properties of the fluid. The principal mode of deformation to which a
material element is subjected as the flow converges into a constriction involves both a shearing of
the material element and a stretching or elongation in the direction of flow, while in the
diverging portion the flow involves both shearing and compression. The actual channel geometry
determines the ratio of shearing to extensional contributions. In many realistic situations
involving \vc\ flows the extensional contribution is the most important of the two modes. As porous
media flow involves elongational flow components, the coil-stretch phenomenon can also take place.
Consequently, a suitable model pore geometry is one having converging and diverging sections which
can reproduce the elongational nature of the flow, a feature that is not exhibited by straight
capillary tubes \cite{MarshallM1967, deiberthesis, denysthesis, DurstHI1987, PilitsisB1989}.

For long time, the straight capillary tube has been the conventional model for porous media and
packed beds. Despite the general success of this model with the \NEW\ and inelastic \nNEW\ flow,
its failure with elastic flow was remarkable. To redress the flaws of this model, the undulating
tube and \convdiv\ channel were proposed in order to include the elastic character of the flow.
Various corrugated tube models with different simple geometries have been used as a first step to
model the effect of \convdiv\ geometry on the flow of \vc\ fluids in porous media (e.g.
\cite{MarshallM1967, DeiberS1981, SouvaliotisB1992}). Those geometries include conically shaped
sections, sinusoidal corrugation and abrupt expansions and contractions. Similarly, a bundle of
\convdiv\ tubes forms a better model for a porous medium in \vc\ flow than the normally used bundle
of straight capillary tubes, as the presence of diameter variations makes it possible to account
for elongational contributions.

Many investigators have attempted to capture the role of the successive \convdiv\ character of
packed bed flow by numerically solving the flow equations in conduits of periodically varying cross
sections. Different opinions on the success of these models can be found in the literature.
Examples include \cite{denysthesis, deiberthesis, TalwarK1992, ChhabraCM2001, PilitsisB1989,
PodolsakTF1997}. With regards to modeling \vc\ flow in regular or random networks of \convdiv\
capillaries, very little work has been done.

\newpage

\section{Modeling the Flow in Porous Media} \label{ModelingFlowPorous}
The basic equations describing the flow of fluids consist of the basic laws of continuum mechanics
which are the conservation principles of mass, energy and linear and angular momentum. These
governing equations indicate how the mass, energy and momentum of the fluid change with position
and time. The basic equations have to be supplemented by a suitable rheological equation of state,
or constitutive equation describing a particular fluid, which is a differential or integral
mathematical relationship that relates the extra stress tensor to the rate-of-strain tensor in
general flow condition and closes the set of governing equations. One then solves the constitutive
model together with the conservation laws using a suitable method to predict velocity and stress
fields of the flows \cite{birdbook, Hulsen1990, Hulsen1986, carreaubook, Keunings2003,
Keunings2004}.

The mathematical description of the flow in porous media is extremely complex task and involves
many approximations. So far, no analytical fluid mechanics solution to the flow through porous
media has been found. Furthermore, such a solution is apparently out of reach for the foreseeable
future. Therefore, to investigate the flow through porous media other methodologies have been
developed, the main ones are the macroscopic continuum approach and the pore-scale numerical
approach.

The widely used continuum models represent a simplified macroscopic approach in which the porous
medium is treated as a continuum. All the complexities and fine details of the microscopic pore
structure are absorbed into bulk terms like permeability that reflect average properties of the
medium. Semi-empirical equations such as \Darcy's law, \Blake-\Kozeny-\Carman\ or \Ergun\ equation
fall into this category. Several continuum models are based in their derivation on the capillary
bundle concept. The advantage of the continuum method is that it is simple and easy to implement
with no computational cost. The disadvantage is that it does not account for the detailed physics
at the pore level. One consequence of this is that in most cases it can only deal with \steadys\
situations with no \timedep\ \trans\ effects.

In the numerical approach, a detailed description of the porous medium at pore-scale level is
adopted and the relevant physics of flow at this level is applied. To find the solution, numerical
methods, such as finite volume and finite difference, usually in conjunction with computational
implementation are used. The advantage of the numerical method is that it is the most direct
approach to describe the physical situation and the closest to full analytical solution. It is also
capable, in principle at least, to deal with \timedep\ \trans\ situations. The disadvantage is that
it requires a detailed pore space description. Moreover, it is usually very complex and hard to
implement and has a huge computational cost with serious convergence difficulties. Due to these
complexities, the flow processes and porous media that are currently within the reach of numerical
investigation are the most simple ones.

Pore-scale network modeling is a relatively novel method developed to deal with flow through porous
media. It can be seen as a compromise between these two extreme approaches as it partly accounts
for the physics and void space description at the pore level with reasonable and generally
affordable computational cost. Network modeling can be used to describe a wide range of properties
from capillary pressure characteristics to interfacial area and mass transfer coefficients. The
void space is described as a network of pores connected by throats. The pores and throats are
assigned some idealized geometry, and rules which determine the transport properties in these
elements are incorporated in the network to compute effective transport properties on a mesoscopic
scale. The appropriate pore-scale physics combined with a geologically representative description
of the pore space gives models that can successfully predict average behavior \cite{blunt2,
blunt1}.

In our investigation to the flow of \BauMan\ fluids in porous media we use network modeling. Our
model uses three-dimensional networks built from a topologically-equivalent three-dimensional voxel
image of the pore space with the pore sizes, shapes and connectivity reflecting the real medium.
Pores and throats are modeled as having triangular, square or circular cross-section by assigning a
shape factor which is the ratio of the area to the perimeter squared and obtained from the pore
space image. Most of the network elements are not circular. To account for the non-circularity when
calculating the volumetric flow rate analytically or numerically for a cylindrical capillary, an
equivalent radius $R_{eq}$ is defined:
\begin{equation}
    \verb|       |    R_{eq} = \left( \frac{8G}{\pi} \right)^{1/4}
\end{equation}
where the geometric conductance, $G$, is obtained empirically from numerical simulation. Two
networks obtained from Statoil and representing two different porous media have been used: a
\sandp\ and a \Berea\ sandstone. These networks are constructed by {\O}ren and coworkers
\cite{OrenBA1997, OrenB2003} from voxel images generated by simulating the geological processes by
which the porous medium was formed. The physical and statistical properties of the networks are
given in \cite{SochiB2008}.

Assuming a laminar, isothermal and incompressible flow at low \Rey\ number, the only equations that
need to be considered are the constitutive equation for the particular fluid and the conservation
of volume as an expression for the conservation of mass. Because initially the pressure drop in
each network element is not known, an iterative method is used. This starts by assigning an
effective viscosity to each network element. The effective viscosity is defined as that viscosity
which makes Poiseulle's equation fit any set of laminar flow conditions for \timeind\ fluids
\cite{skellandbook}. By invoking the conservation of volume for incompressible fluid, the pressure
field across the entire network is solved using a numerical solver \cite{rugebook}. Knowing the
pressure drops, the effective viscosity of each element is updated using the expression for the
flow rate with a pseudo-\POIS\ definition. The pressure field is then recomputed using the updated
viscosities and the iteration continues until convergence is achieved when a specified error
tolerance in total flow rate between two consecutive iteration cycles is reached. Finally, the
total volumetric flow rate and the apparent viscosity in porous media, defined as the viscosity
calculated from the \Darcy's law, are obtained.

With regards to modeling the flow in porous media of complex fluids which have time dependency due
to \thixotropic\ or elastic nature, there are three major difficulties :
\begin{itemize}

\item The difficulty of tracking the fluid elements in the pores and throats
and identifying their deformation history, as the path followed by these elements is random and can
have several unpredictable outcomes.

\item The mixing of fluid elements with various deformation history in
the individual pores and throats. As a result, the viscosity is not a well-defined property of the
fluid in the pores and throats.

\item The change of viscosity along the streamline since the deformation history is
constantly changing over the path of each fluid element.

\end{itemize}
In the current work, we deal only with one case of \steadys\ \vc\ flow, and hence we did not
consider these complications in depth. Consequently, the tracking of fluid elements or flow history
in the network and other dynamic aspects are not implemented in the \nNEW\ code. However, \timedep\
effects in \steadys\ conditions are accounted for in the \Tardy\ algorithm which is implemented in
the code and will be presented in detail in \S\ (\ref{TardyAlgorithm}).

In our modeling approach, to solve the pressure field across a network of $n$ nodes we write $n$
equations in $n$ unknowns which are the pressure values at the nodes. The essence of these
equations is the continuity of flow of incompressible fluid at each node in the absence of source
and sink. We solve this set of equations subject to the boundary conditions which are the pressures
at the inlet and outlet. This unique solution is `consistent' and `stable' as it is the only
mathematically acceptable solution to the problem, and, assuming the modeling process and the
mathematical technicalities are correct, should mimic the unique physical reality of the pressure
field in the porous medium.

\newpage

\section{\BauMan\ Model} \label{BautistaManero}

This is a relatively simple model that combines the \OldB\ constitutive equation for \vy\ and the
\FRED's kinetic equation for flow-induced structural changes usually associated with \thixotropy.
The model requires six parameters that have physical significance and can be estimated from
rheological measurements. These parameters are the low and high shear rate viscosities, the elastic
modulus, the relaxation time, and two other constants describing the build up and break down of
viscosity.

The \OldB\ model is a simplification of the more elaborate and rarely used \OLD\ 8-constant model
which also contains the upper convected, the lower convected, and the corotational \Maxwell\
equations as special cases. \OldB\ is the second simplest nonlinear \vc\ model and is apparently
the most popular in \vc\ flow modeling and simulation. It is the nonlinear equivalent of the linear
\JEF\ model, and hence it takes account of frame invariance in the nonlinear regime. Consequently,
in the linear \vc\ regime the \OldB\ model reduces to the linear \JEF\ model. The \OldB\ model can
be obtained by replacing the partial time derivatives in the differential form of the \JEF\ model
with the upper convected time derivatives \cite{birdbook}
\begin{equation}\label{OBM1}
    {\sTen} + \rxTim {\ucd \sTen} =
    \lVis \left( {\rsTen} + \rdTim {\ucd \rsTen} \right)
\end{equation}
where $\sTen$ is the extra stress tensor, $\rxTim$ is the relaxation time, $\rdTim$ is the
retardation time, $\lVis$ is the low-shear viscosity, $\rsTen$ is the rate-of-strain tensor, and
{$\ucd \sTen$} is the upper convected time derivative of the stress tensor:

\begin{equation}\label{UCM2}
    \ucd \sTen =
    \frac{\partial {\sTen}}{\partial t} +
    \fVel \cdot \nabla \sTen -
    \left( \nabla \fVel \right)^{T} \cdot \sTen -
    \sTen \cdot \nabla \fVel
\end{equation}
where $t$ is time, $\fVel = (\vC_{x}, \vC_{y}, \vC_{z})$ is the fluid velocity vector, $\left(
\cdot \right)^{T}$ is the transpose of the tensor and $\nabla \fVel$ is the fluid velocity gradient
tensor defined by
\begin{equation}\label{fVelGradTen}
    \nabla \fVel =
    \left(
    \begin{array}{ccc}
      \frac{\partial \vC_{x}}{\partial x} &
      \frac{\partial \vC_{x}}{\partial y} &
      \frac{\partial \vC_{x}}{\partial z}
      \\
      \frac{\partial \vC_{y}}{\partial x} &
      \frac{\partial \vC_{y}}{\partial y} &
      \frac{\partial \vC_{y}}{\partial z}
      \\
      \frac{\partial \vC_{z}}{\partial x} &
      \frac{\partial \vC_{z}}{\partial y} &
      \frac{\partial \vC_{z}}{\partial z}
    \end{array}
    \right)
\end{equation}

Similarly, {$\ucd \rsTen$} is the upper convected time derivative of the rate-of-strain tensor
given by:
\begin{equation}\label{OBM2}
    \ucd \rsTen =
    \frac{\partial {\rsTen}}{\partial t} +
    \fVel \cdot \nabla \rsTen -
    \left( \nabla \fVel \right)^{T} \cdot \rsTen -
    \rsTen \cdot \nabla \fVel
\end{equation}

The kinetic equation of \FRED\ that accounts for the destruction and construction of structure is
given by
\begin{equation}\label{Fredrickson}
    \D \Vis t = \frac{\Vis}{\rxTimF} \left( 1 - \frac{\Vis}{\lVis} \right)
                + \kF \Vis \left( 1 - \frac{\Vis}{\hVis} \right) \sTen : \rsTen
\end{equation}
where  $\Vis$ is the \nNEW\ viscosity, $t$ is the time of deformation, $\rxTimF$ is the relaxation
time upon the cessation of steady flow, $\lVis$ and $\hVis$ are the viscosities at zero and
infinite shear rates respectively, $\kF$ is a parameter that is related to a critical stress value
below which the material exhibits primary creep, $\sTen$ is the stress tensor and $\rsTen$ is the
rate of strain tensor. In this model, $\rxTimF$ is a structural relaxation time, whereas $\kF$ is a
kinetic constant for structure break down. The elastic modulus $\Go$ is related to these parameters
by $\Go = \Vis/\rxTimF$ \cite{BautistaSPM1999, BautistaSLPM2000, ManeroBSP2002, TardyA2005}.

\BauMan\ model was originally proposed for the rheology of worm-like micellar solutions which
usually have an upper \NEW\ plateau, and show strong signs of \shThin. The model, which
incorporates \shThin, elasticity and \thixotropy, can be used to describe the complex rheological
behavior of \vc\ systems that also exhibit \thixotropy\ and \rheopexy\ under shear flow. The model
predicts creep behavior, stress relaxation and the presence of \thixotropic\ loops when the sample
is subjected to \trans\ stress cycles. The \BauMan\ model has also been found to fit steady shear,
oscillatory and \trans\ measurements of \vc\ solutions \cite{BautistaSLPM2000, TardyA2005,
BautistaSPM1999, ManeroBSP2002}.

\subsection{\Tardy\ Algorithm}\label{TardyAlgorithm}
This algorithm is proposed by Philippe \Tardy\ to compute the pressure drop-flow rate relationship
for the \steadys\ flow of a \BauMan\ fluid in simple capillary network models. The bulk rheology of
the fluid and the dimensions of the capillaries making up the network are used as inputs to the
models \cite{TardyA2005}. In the following paragraphs we outline the basic components of this
algorithm and the logic behind it. This will be followed by some mathematical and technical details
related to the implementation of this algorithm in our \nNEW\ code.

The flow in a single capillary can be described by the following general relation
\begin{equation}\label{generalFlowPresRel}
    Q = G' \Delta P
\end{equation}
where $Q$ is the volumetric flow rate, $G'$ is the flow conductance and $\Delta P$ is the pressure
drop. For a particular capillary and a specific fluid, $G'$ is given by
\begin{eqnarray}
  G' &=& G'(\mu) = \textrm{constant}                     \verb|      |   \textrm{\NEW\ Fluid} \nonumber \\
  G' &=& G'(\mu, \Delta P)            \verb|            |      \textrm{Purely viscous \nNEW\ Fluid} \nonumber \\
  G' &=& G'(\mu, \Delta P, t)         \verb|           |      \textrm{Fluid with memory}
\end{eqnarray}

For a network of capillaries, a set of equations representing the capillaries and satisfying mass
conservation should be solved simultaneously to produce a consistent pressure field, as presented
in \S\ (\ref{ModelingFlowPorous}). For \NEW\ fluid, a single iteration is needed to solve the
pressure field since the conductance is known in advance as the viscosity is constant. For purely
viscous \nNEW\ fluid, we start with an initial guess for the viscosity, as it is unknown and
pressure-dependent, and solve the pressure field iteratively updating the viscosity after each
iteration cycle until convergence is reached. For memory fluids, the dependence on time must be
taken into account when solving the pressure field iteratively. Apparently, there is no general
strategy to deal with such situation. However, for the \steadys\ flow of memory fluids a sensible
approach is to start with an initial guess for the flow rate and iterate, considering the effect of
the local pressure and viscosity variation due to \convdiv\ geometry, until convergence is
achieved. This approach is adopted by \Tardy\ to find the flow of a \BauMan\ fluid in a simple
capillary network model. In general terms, the \Tardy\ algorithm can be summarized as follows
\cite{TardyA2005}
\begin{itemize}

\item For fluids without memory the capillary is unambiguously defined by its radius and length. For
fluids with memory, where going from one section to another with different radius is important, the
capillary should be modeled with contraction to account for the effect of \convdiv\ geometry on the
flow. The reason is that the effects of fluid memory take place on going through a radius change,
as this change induces a change in shear rate and generation of extensional flow fields with a
consequent \vc\ and \thixotropic\ effects. Examples of the \convdiv\ geometries are given in Figure
(\ref{ConvDivGeom}).

\item Each capillary is discretized in the flow direction and a discretized form
of the flow equations is used assuming a prior knowledge of the stress and viscosity at the inlet
of the network.

\item Starting with an initial guess for the flow rate and using iterative technique,
the pressure drop as a function of the flow rate is found for each capillary.

\item Finally, the pressure field for the whole network is found iteratively
until convergence is achieved. Once the pressure field is found the flow rate through each
capillary in the network can be computed and the total flow rate through the network can be
determined by summing and averaging the flow through the inlet and outlet capillaries.

\end{itemize}

A modified version of the \Tardy\ algorithm was implemented in our \nNEW\ code. In the following,
we outline the main steps of this algorithm and its implementation.

\begin{itemize}

\item From a one-dimensional \steadys\ \FRED\ equation in which the partial time derivative is written in
the form $\Dp{}{t} = V \Dp{}{x}$, the following equation can be obtained

\begin{equation}\label{simplifiedFred}
    V \D \Vis x = \frac{\Vis}{\rxTimF} \left( \frac{\lVis - \Vis}{\lVis} \right)
    + \kF \Vis \left( \frac{\hVis - \Vis}{\hVis} \right) \sTenC \rsTenC
\end{equation}

In this formulation, the fluid speed $V$ is given by $Q / \pi r^{2}$ and the average shear rate in
the tube with radius $r$ is given by $Q / \pi r^{3}$.

\item By similar argument, another simplified equation can be obtained from
the \OldB\ model

\begin{equation}\label{simplifiedOldB}
    \sTenC + \frac{V \Vis}{\Go} \D \sTenC x = 2 \Vis \rsTenC
\end{equation}

\item The \convdiv\ feature of the capillary is implemented in the form of a parabolic profile as
outlined in Appendix A.

\item Each capillary in the network is discretized into $m$ slices, each with width $\delta x = L/m$
where $L$ is the capillary length.

\item From Equation (\ref{simplifiedOldB}), applying the simplified assumption $\D \sTenC x =
\frac{(\sTenC_{2} - \sTenC_{1})}{\delta x}$ where the subscripts $1$ and $2$ stand for the inlet
and outlet of the slice respectively, we obtain an expression for $\sTenC_{2}$ in terms of
$\sTenC_{1}$ and $\Vis_{2}$

\begin{equation}\label{tau2}
    \sTenC_{2} = \frac{2 \Vis_{2} \rsTenC + \frac{V \Vis_{2} \sTenC_{1}}{\Go \delta x}}
    {1 + \frac{V \Vis_{2}}{\Go \delta x}}
\end{equation}

\item From Equations (\ref{simplifiedFred}) and (\ref{tau2}), applying $\D \Vis x = \frac{(\Vis_{2} -
\Vis_{1})}{\delta x}$, a third order polynomial in $\Vis_{2}$ is obtained. The coefficients of the
four terms of this polynomial are
\begin{eqnarray} \label{mu2}
  \Vis_{2}^{3}: \hspace{1.0cm} - \frac{V}{\rxTimF \lVis \Go \delta x} - \frac{2 \kF \rsTenC^{2}}{\hVis} - \frac{\kF \rsTenC \sTenC_{1} V}{\hVis \Go \delta x} \hspace{2.4cm} \nonumber \\
  \Vis_{2}^{2}: \hspace{1.0cm} - \frac{1}{\rxTimF \lVis} - \frac{V^{2}}{\Go (\delta x)^{2}} + \frac{V}{\rxTimF \Go \delta x} + 2 \kF \rsTenC^{2} + \frac{\kF \rsTenC \sTenC_{1} V}{\Go \delta x} \nonumber \\
  \Vis_{2}^{1}: \hspace{1.0cm} - \frac{V}{\delta x} + \frac{1}{\rxTimF} + \frac{V^{2} \Vis_{1}}{\Go (\delta x)^{2}} \hspace{4.2cm} \nonumber \\
  \Vis_{2}^{0}: \hspace{1.0cm} \frac{V \Vis_{1}}{\delta x} \hspace{7.1cm}
\end{eqnarray}

\item The algorithm starts by assuming a \NEW\ flow in a network of straight capillaries. Accordingly,
the volumetric flow rate $Q$, and consequently $V$ and $\rsTenC$, for each capillary are obtained.

\item Starting from the inlet of the network where the viscosity and the stress are assumed to have known
values of $\lVis$ and $R \Delta P / 2 L$ for each capillary respectively, the computing of the
\nNEW\ flow in a \convdiv\ geometry takes place in each capillary independently by calculating
$\Vis_{2}$ and $\sTenC_{2}$ slice by slice, where the values from the previous slice are used for
$\Vis_{1}$ and $\sTenC_{1}$ of the current slice.

\item For the capillaries which are not at the inlet of the network, the initial values of the viscosity
and stress at the inlet of the capillary are found by computing the $Q$-weighted average from all
the capillaries that feed into the pore which is at the inlet of the corresponding capillary.

\item To find $\Vis_{2}$ of a slice, a bisection numerical method is used. To eliminate possible
non-physical roots, the interval for the accepted root is set between zero and $3\lVis$ with
$\lVis$ used in the case of failure. These conditions are logical as long as the slice is
reasonably thin and the flow and fluid are physically viable. In the case of a convergence failure,
error messages are issued to inform the user. No failure has been detected during the many runs of
this algorithm. Moreover, extensive sample inspection of the $\Vis_{2}$ values has been carried out
and proved to be sensible and realistic.

\item The value found for the $\Vis_{2}$ is used in conjunction with Equation (\ref{tau2}) to find
$\sTenC_{2}$ which is needed as an input to the next slice.

\item Averaging the value of $\Vis_{1}$ and $\Vis_{2}$, the viscosity for the slice is found and used
with \POIS\ law to find the pressure drop across the slice.

\item The total pressure drop across the whole capillary is computed by summing up the pressure drops
across its individual slices. This total is used with \POIS\ law to find the effective viscosity
for the capillary as a whole.

\item Knowing the effective viscosities for all capillaries of the network, the pressure field is solved
iteratively using an algebraic multi-grid  solver, and hence the total volumetric flow rate from
the network and the apparent viscosity are found.

\end{itemize}

\subsection{Initial Results of the Modified \Tardy\ Algorithm}\label{ResultsTardyAlgorithm}
The modified \Tardy\ algorithm was tested and assessed. Various qualitative aspects were verified
and proved to be correct. Some general results and conclusions are outlined below with sample
graphs and data for the \sandp\ and \Berea\ networks using a calculation box with $x_{_{l}}$=0.5
and $x_{_{u}}$=0.95. We would like to remark that despite our effort to use typical values for the
parameters and variables, in some cases we were forced, for demonstration purposes, to use
eccentric values to accentuate the features of interest. In Table (\ref{wormlikeMicellar}) we
present some values for the \BauMan\ model parameters as obtained from the wormlike micellar system
studied by Anderson and co-workers \cite{AndersonPS2006} to give an idea of the parameter ranges in
real fluids. The system is a solution of a surfactant concentrate [a mixture of the cationic
surfactant erucyl bis(hydroxyethyl)methylammonium chloride (EHAC) and 2-propanol in a 3:2 weight
ratio] in an aqueous solution of potassium chloride.


\begin{table} [h]
\centering %
\caption[Some values of the wormlike micellar system studied by Anderson and co-workers, which is a
solution of surfactant concentrate (a mixture of EHAC and 2-propanol) in an
aqueous solution of potassium chloride] %
{Some values of the wormlike micellar system studied by Anderson and co-workers
\cite{AndersonPS2006} which is a solution of surfactant concentrate (a mixture of EHAC and
2-propanol) in an
aqueous solution of potassium chloride.} %
\label{wormlikeMicellar} %
\vspace{0.5cm} %
\begin{tabular}{|l|l|}
\hline

Parameter   \verb|  |  &           Value \verb|  | \\

\hline

$\Go$ (Pa)              &          1 - 10 \\

$\lVis$ (Pa.s)          &          115 - 125 \\

$\hVis$ (Pa.s)          &          0.00125 - 0.00135 \\

$\rxTimF$ (s)           &          1 - 30 \\

$\kF$ (Pa$^{-1}$)       &          10$^{^{-3}}$ - 10$^{^{-6}}$ \\

\hline
\end{tabular}
\end{table}


\subsubsection{Convergence-Divergence}\label{}
A dilatant effect due to the \convdiv\ feature has been detected relative to a network of straight
capillaries. The viscosity increase is a natural viscoelastic response to the radius tightening at
the middle. As the corrugation feature of the tubes is exacerbated by narrowing the radius at the
middle, the \vc\ dilatant behavior is intensified. The natural explanation is that the increase in
apparent viscosity should be proportionate to the magnitude of tightening. This feature is
presented in Figures (\ref{TardySP1}) and (\ref{TardyB1}) for the \sandp\ and \Berea\ networks
respectively on a log-log scale.

\begin{figure}[!t]
  \centering{}
  \includegraphics
  [scale=0.5]
  {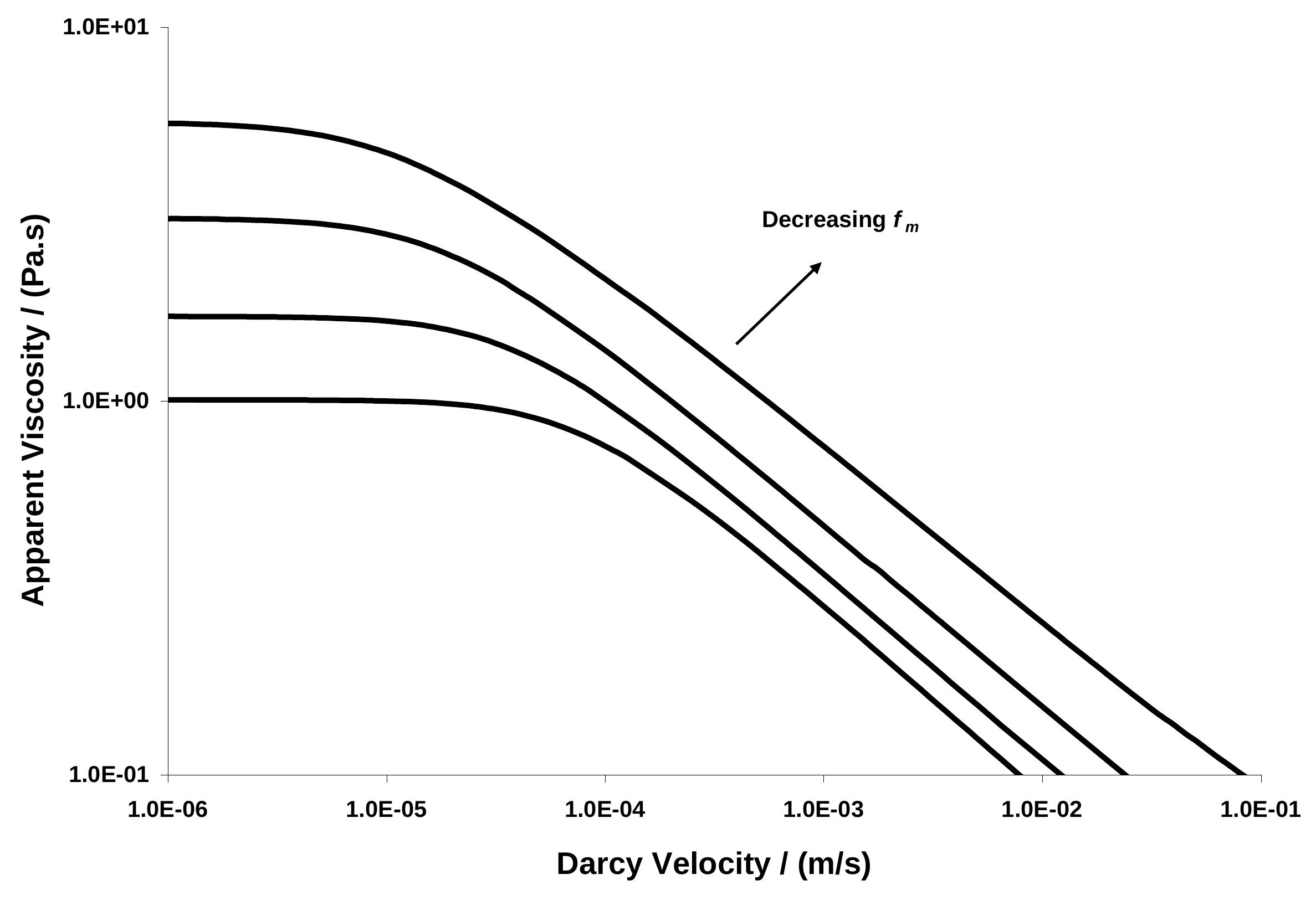}
  \caption[The \Tardy\ algorithm \sandp\ results for $\Go$=0.1\,Pa, $\hVis$=0.001\,Pa.s, $\lVis$=1.0\,Pa.s, $\rxTimF$=1.0\,s,
            $\kF$=10$^{-5}$\,Pa$^{-1}$, $\fe$=1.0, $m$=10 slices, with varying $\fm$ (1.0, 0.8, 0.6 and 0.4)]
  {The \Tardy\ algorithm \sandp\ results for $\Go$=0.1\,Pa, $\hVis$=0.001\,Pa.s, $\lVis$=1.0\,Pa.s, $\rxTimF$=1.0\,s,
            $\kF$=10$^{-5}$\,Pa$^{-1}$, $\fe$=1.0, $m$=10 slices, with varying $\fm$ (1.0, 0.8, 0.6 and 0.4).}
  \label{TardySP1}
\end{figure}
\begin{figure}[!h]
  \centering{}
  \includegraphics
  [scale=0.5]
  {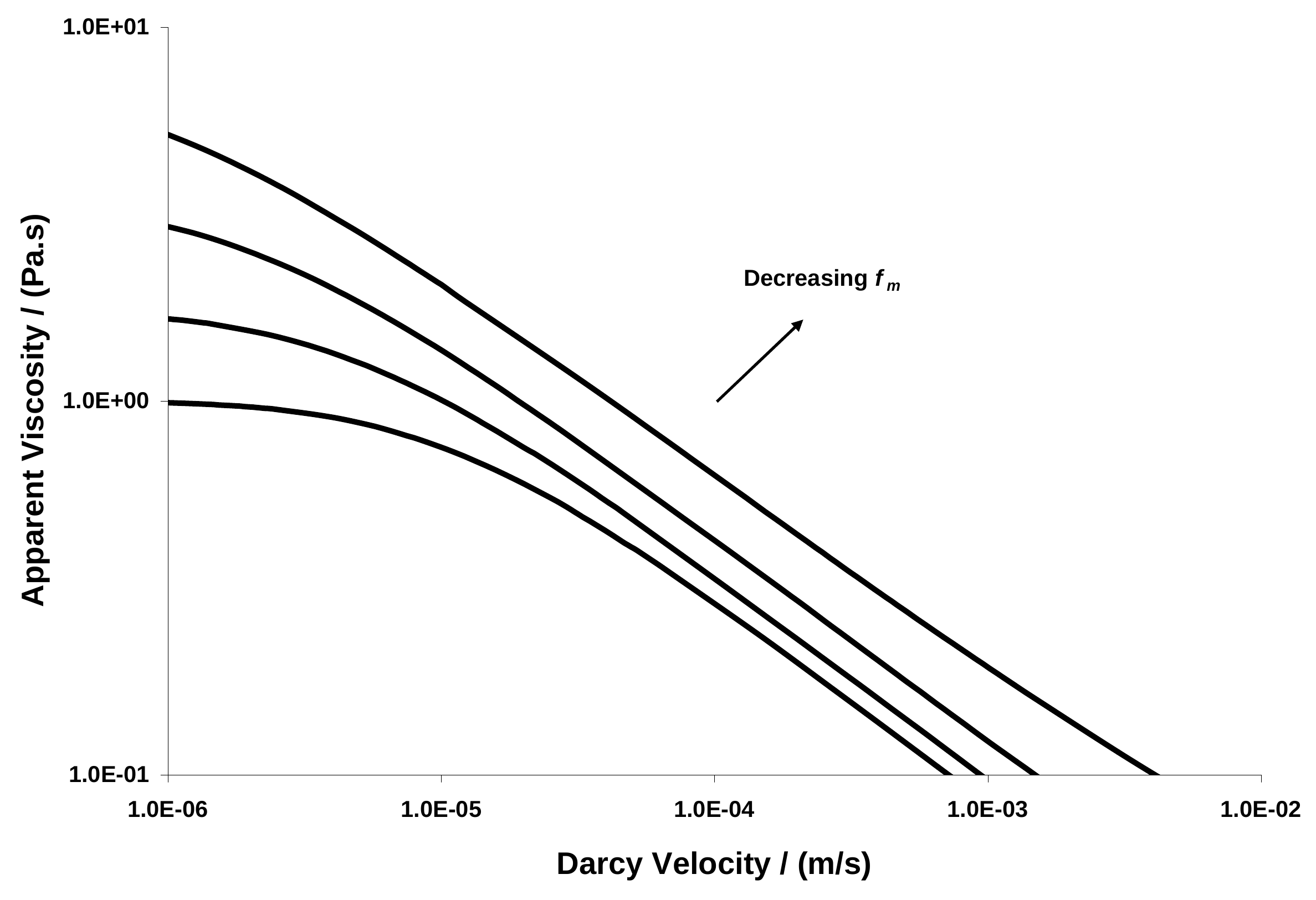}
  \caption[The \Tardy\ algorithm \Berea\ results for $\Go$=0.1\,Pa, $\hVis$=0.001\,Pa.s, $\lVis$=1.0\,Pa.s, $\rxTimF$=1.0\,s,
            $\kF$=10$^{-5}$\,Pa$^{-1}$, $\fe$=1.0, $m$=10 slices, with varying $\fm$ (1.0, 0.8, 0.6 and 0.4)]
  {The \Tardy\ algorithm \Berea\ results for $\Go$=0.1\,Pa, $\hVis$=0.001\,Pa.s, $\lVis$=1.0\,Pa.s, $\rxTimF$=1.0\,s,
            $\kF$=10$^{-5}$\,Pa$^{-1}$, $\fe$=1.0, $m$=10 slices, with varying $\fm$ (1.0, 0.8, 0.6 and 0.4).}
  \label{TardyB1}
\end{figure}

\subsubsection{\DivConv}\label{}
The investigation of the effect of \divconv\ geometry by expanding the radius of the capillaries at
the middle revealed a thinning effect relative to the straight and \convdiv\ geometries, as seen in
Figures (\ref{TardySP2}) and (\ref{TardyB2}) for the \sandp\ and \Berea\ networks respectively on a
log-log scale. This is due to \vc\ response in the opposite sense to that of \convdiv\ by enlarging
the radius at the middle.

\begin{figure}[!t]
  \centering{}
  \includegraphics
  [scale=0.5]
  {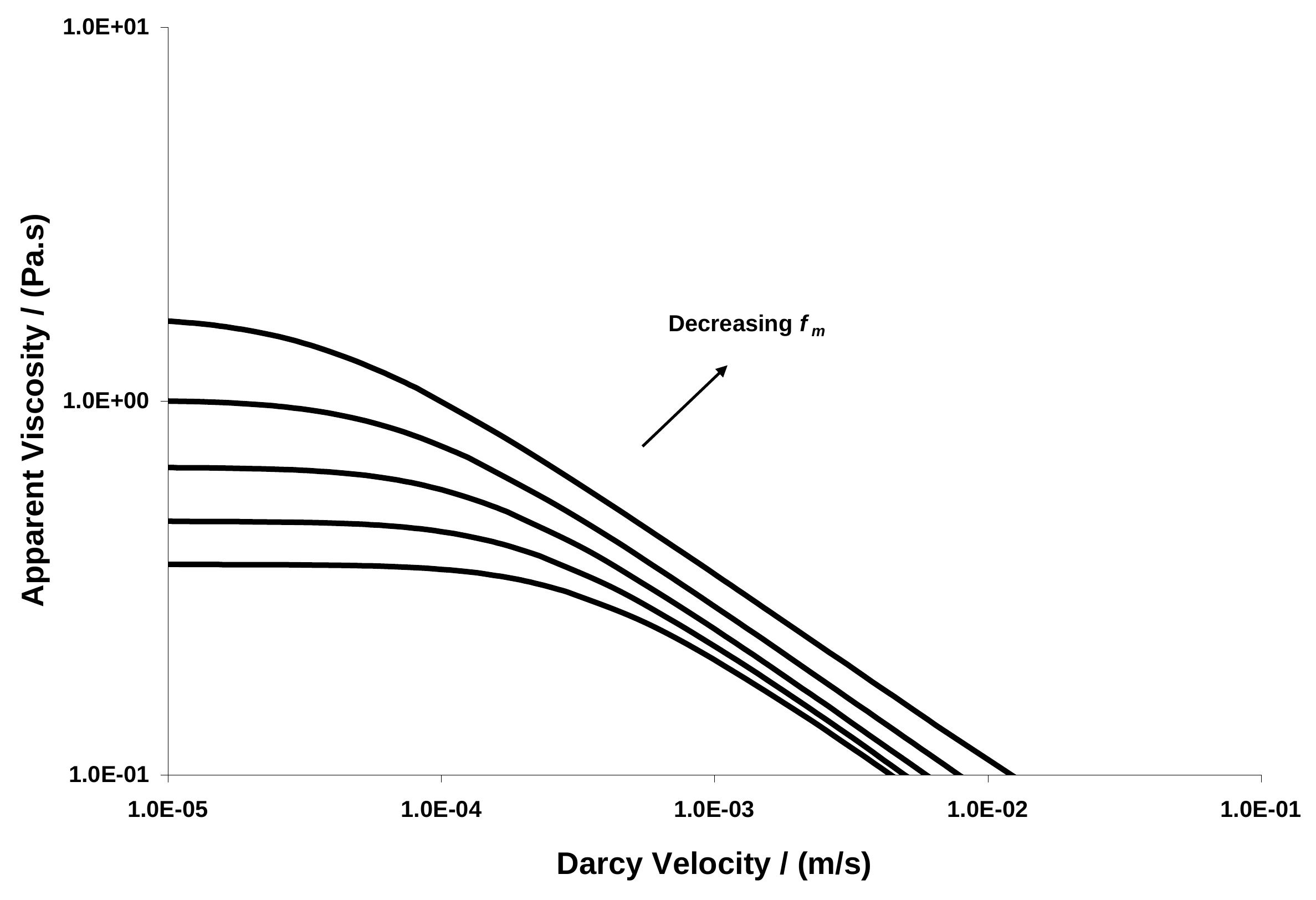}
  \caption[The \Tardy\ algorithm \sandp\ results for $\Go$=0.1\,Pa, $\hVis$=0.001\,Pa.s, $\lVis$=1.0\,Pa.s, $\rxTimF$=1.0\,s,
            $\kF$=10$^{-5}$\,Pa$^{-1}$, $\fe$=1.0, $m$=10 slices, with varying $\fm$ (0.8, 1.0, 1.2, 1.4 and 1.6)]
  {The \Tardy\ algorithm \sandp\ results for $\Go$=0.1\,Pa, $\hVis$=0.001\,Pa.s, $\lVis$=1.0\,Pa.s, $\rxTimF$=1.0\,s,
            $\kF$=10$^{-5}$\,Pa$^{-1}$, $\fe$=1.0, $m$=10 slices, with varying $\fm$ (0.8, 1.0, 1.2, 1.4 and 1.6).}
  \label{TardySP2}
\end{figure}
\begin{figure}[!h]
  \centering{}
  \includegraphics
  [scale=0.5]
  {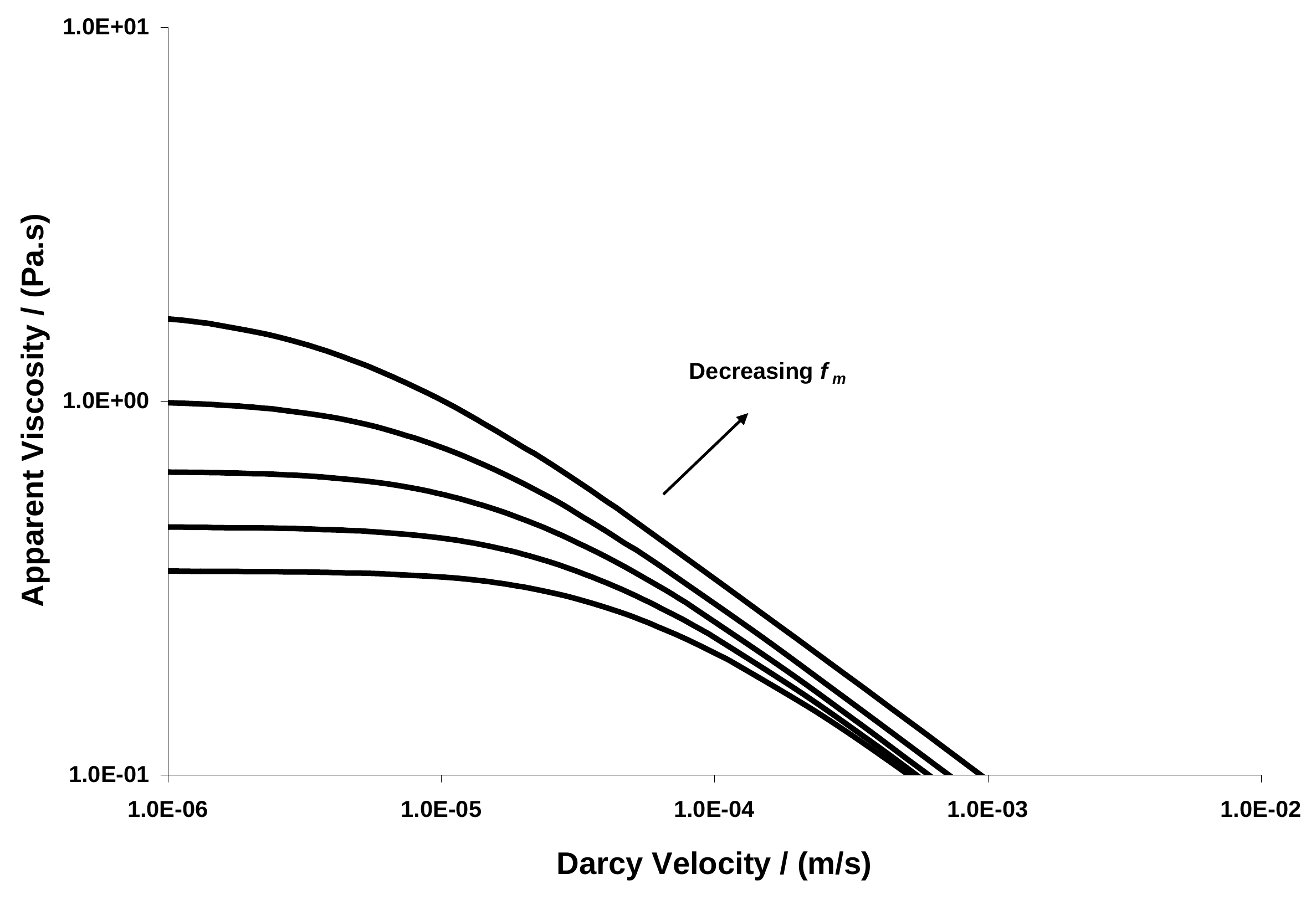}
  \caption[The \Tardy\ algorithm \Berea\ results for $\Go$=0.1\,Pa, $\hVis$=0.001\,Pa.s, $\lVis$=1.0\,Pa.s, $\rxTimF$=1.0\,s,
            $\kF$=10$^{-5}$\,Pa$^{-1}$, $\fe$=1.0, $m$=10 slices, with varying $\fm$ (0.8, 1.0, 1.2, 1.4 and 1.6)]
  {The \Tardy\ algorithm \Berea\ results for $\Go$=0.1\,Pa, $\hVis$=0.001\,Pa.s, $\lVis$=1.0\,Pa.s, $\rxTimF$=1.0\,s,
            $\kF$=10$^{-5}$\,Pa$^{-1}$, $\fe$=1.0, $m$=10 slices, with varying $\fm$ (0.8, 1.0, 1.2, 1.4 and 1.6).}
  \label{TardyB2}
\end{figure}

\subsubsection{Number of Slices}\label{}
As the number of slices of the capillaries increases, the algorithm converges to a stable and
constant solution within acceptable numerical errors. This indicates that the numerical aspects of
the algorithm are functioning correctly because the effect of discretization  errors is expected to
diminish by increasing the number of slices and hence decreasing their width. A sample graph of
apparent viscosity versus number of slices for a typical data point is presented in Figure
(\ref{TardySP3}) for the \sandp\ and in Figure (\ref{TardyB3}) for \Berea.

\begin{figure}[!t]
  \centering{}
  \includegraphics
  [scale=0.5]
  {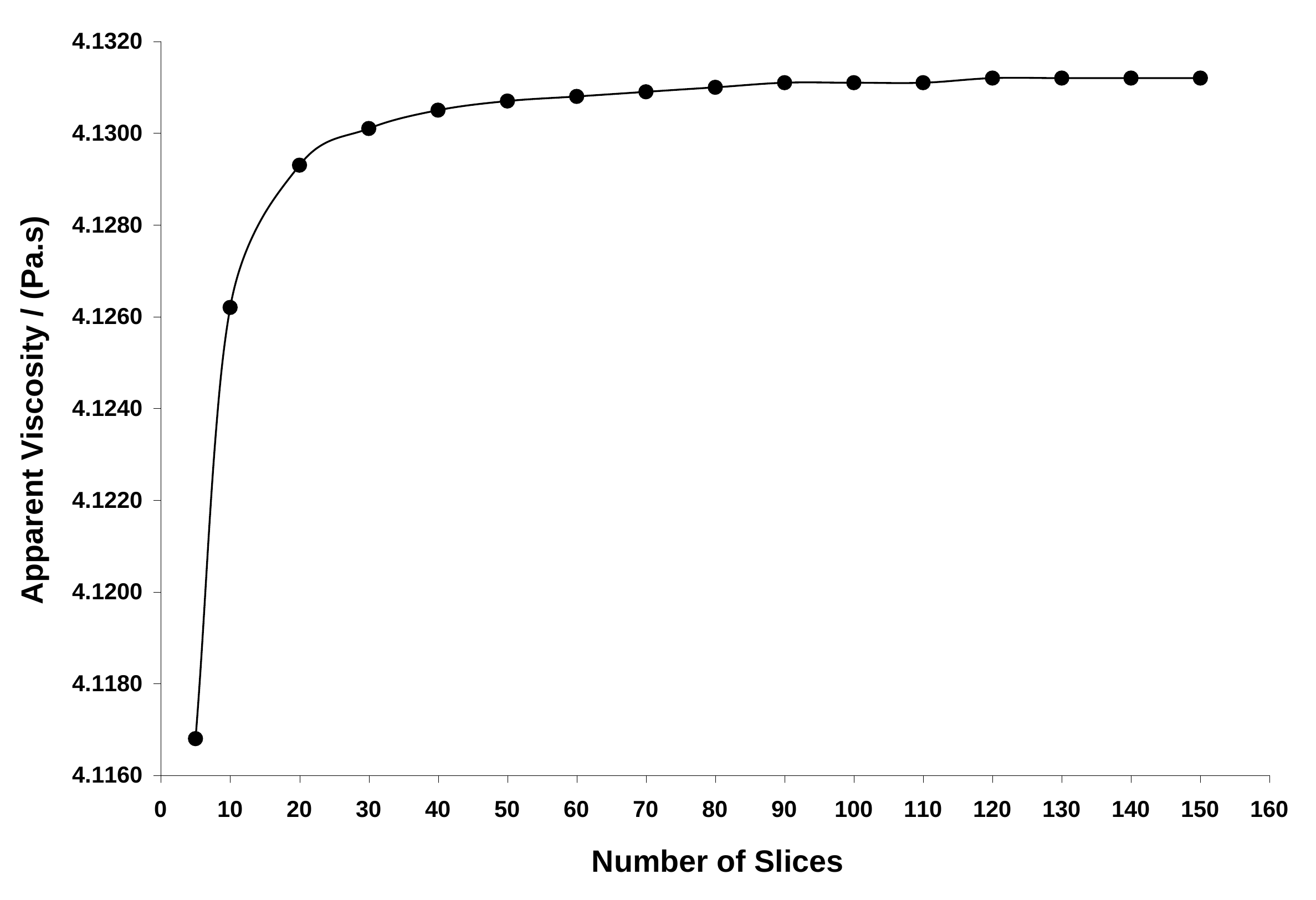}
  \caption[The \Tardy\ algorithm \sandp\ results for $\Go$=1.0\,Pa, $\hVis$=0.001\,Pa.s, $\lVis$=1.0\,Pa.s, $\rxTimF$=1.0\,s,
            $\kF$=10$^{-5}$\,Pa$^{-1}$, $\fe$=1.0, $\fm$=0.5, with varying number of slices for
            a typical data point ($\Delta P$=100\,Pa)]
  {The \Tardy\ algorithm \sandp\ results for $\Go$=1.0\,Pa, $\hVis$=0.001\,Pa.s, $\lVis$=1.0\,Pa.s, $\rxTimF$=1.0\,s,
            $\kF$=10$^{-5}$\,Pa$^{-1}$, $\fe$=1.0, $\fm$=0.5, with varying number of slices for
            a typical data point ($\Delta P$=100\,Pa).}
  \label{TardySP3}
\end{figure}
\begin{figure}[!h]
  \centering{}
  \includegraphics
  [scale=0.5]
  {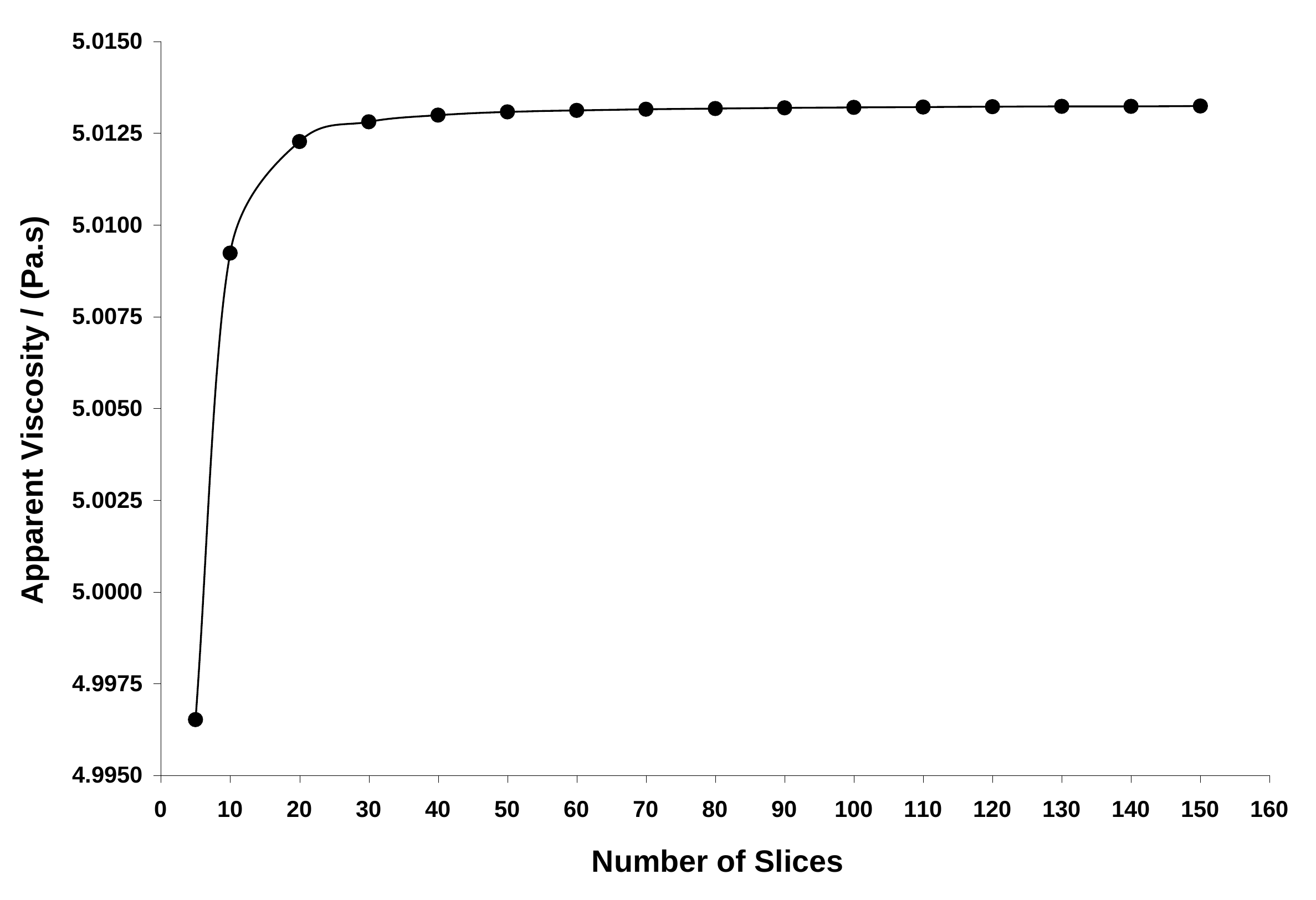}
  \caption[The \Tardy\ algorithm \Berea\ results for $\Go$=1.0\,Pa, $\hVis$=0.001\,Pa.s, $\lVis$=1.0\,Pa.s, $\rxTimF$=1.0\,s,
            $\kF$=10$^{-5}$\,Pa$^{-1}$, $\fe$=1.0, $\fm$=0.5, with varying number of slices for
            a typical data point ($\Delta P$=200\,Pa)]
  {The \Tardy\ algorithm \Berea\ results for $\Go$=1.0\,Pa, $\hVis$=0.001\,Pa.s, $\lVis$=1.0\,Pa.s, $\rxTimF$=1.0\,s,
            $\kF$=10$^{-5}$\,Pa$^{-1}$, $\fe$=1.0, $\fm$=0.5, with varying number of slices for
            a typical data point ($\Delta P$=200\,Pa).}
  \label{TardyB3}
\end{figure}

\subsubsection{\Boger\ Fluid}\label{}
A \Boger\ fluid behavior was observed when setting $\lVis = \hVis$. However, the apparent viscosity
increased as the \convdiv\ feature is intensified. This feature is demonstrated in Figure
(\ref{TardySP4}) for the \sandp\ and in Figure (\ref{TardyB4}) for \Berea\ on a log-log scale.
Since \Boger\ fluid is a limiting and obvious case, this behavior indicates that the model, as
implemented, is well-behaved. The viscosity increase is a natural \vc\ response to the radius
tightening at the middle, as discussed earlier.

\begin{figure}[!t]
  \centering{}
  \includegraphics
  [scale=0.5]
  {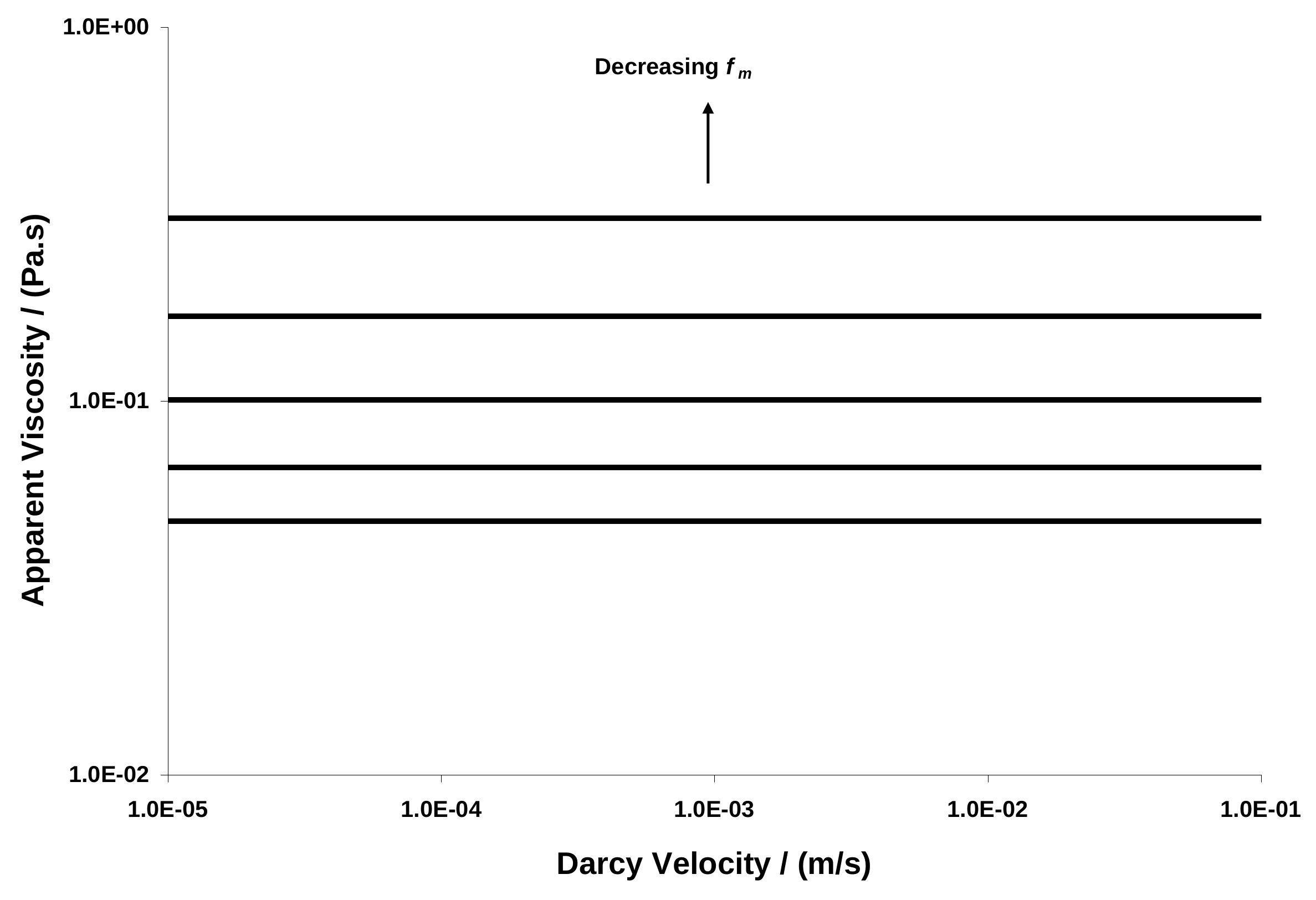}
  \caption[The \Tardy\ algorithm \sandp\ results for $\Go$=1.0\,Pa, $\hVis$=$\lVis$=0.1\,Pa.s, $\rxTimF$=1.0\,s,
            $\kF$=10$^{-5}$\,Pa$^{-1}$, $\fe$=1.0, $m$=10 slices, with varying $\fm$ (0.6, 0.8, 1.0, 1.2 and 1.4)]
  {The \Tardy\ algorithm \sandp\ results for $\Go$=1.0\,Pa, $\hVis$=$\lVis$=0.1\,Pa.s, $\rxTimF$=1.0\,s,
            $\kF$=10$^{-5}$\,Pa$^{-1}$, $\fe$=1.0, $m$=10 slices, with varying $\fm$ (0.6, 0.8, 1.0, 1.2 and 1.4).}
  \label{TardySP4}
\end{figure}
\begin{figure}[!h]
  \centering{}
  \includegraphics
  [scale=0.5]
  {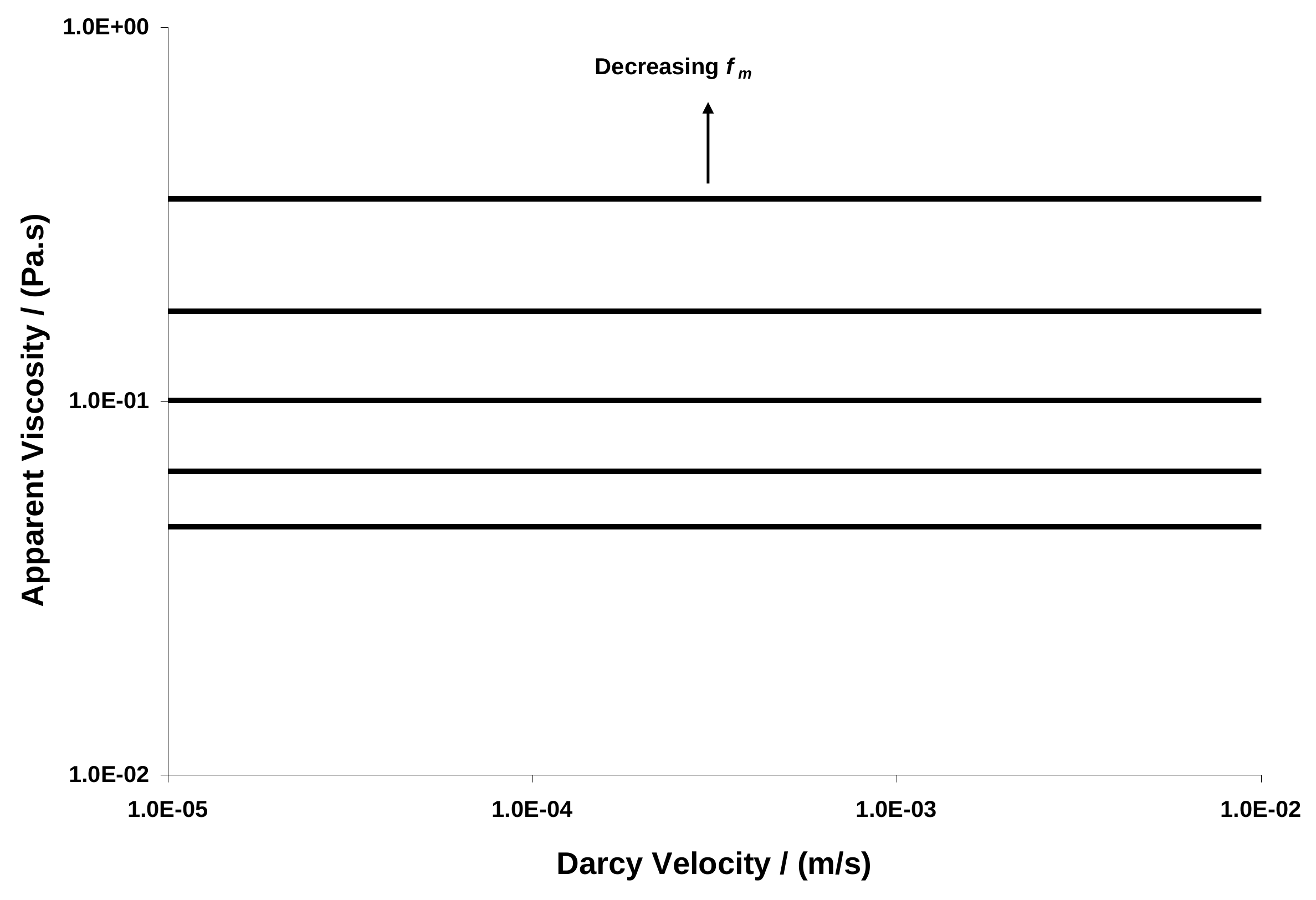}
  \caption[The \Tardy\ algorithm \Berea\ results for $\Go$=1.0\,Pa, $\hVis$=$\lVis$=0.1\,Pa.s, $\rxTimF$=1.0\,s,
            $\kF$=10$^{-5}$\,Pa$^{-1}$, $\fe$=1.0, $m$=10 slices, with varying $\fm$ (0.6, 0.8, 1.0, 1.2 and 1.4)]
  {The \Tardy\ algorithm \Berea\ results for $\Go$=1.0\,Pa, $\hVis$=$\lVis$=0.1\,Pa.s, $\rxTimF$=1.0\,s,
            $\kF$=10$^{-5}$\,Pa$^{-1}$, $\fe$=1.0, $m$=10 slices, with varying $\fm$ (0.6, 0.8, 1.0, 1.2 and 1.4).}
  \label{TardyB4}
\end{figure}

\subsubsection{Elastic Modulus}\label{}
The effect of the elastic modulus $\Go$ was investigated for \shThin\ fluids, i.e. $\lVis > \hVis$,
by varying this parameter over several orders of magnitude while holding the others constant. It
was observed that by increasing $\Go$, the high-shear apparent viscosities were increased while the
low-shear viscosities remained constant and have not been affected. However, the increase at
high-shear rates has almost reached a saturation point where beyond some limit the apparent
viscosities converged to certain values despite a large increase in $\Go$. A sample of the results
for this investigation is presented in Figure (\ref{TardySP5}) for the \sandp\ and in Figure
(\ref{TardyB5}) for \Berea\ on a log-log scale. The stability at low-shear rates is to be expected
because at low flow rate regimes near the lower \NEW\ plateau the \nNEW\ effects due to elastic
modulus are negligible. As the elastic modulus increases, an increase in apparent viscosity due to
elastic effects contributed by elastic modulus occurs. Eventually, a saturation will be reached
when the contribution of this factor is controlled by other dominant factors and mechanisms from
the porous medium and flow regime.

\begin{figure}[!t]
  \centering{}
  \includegraphics
  [scale=0.5]
  {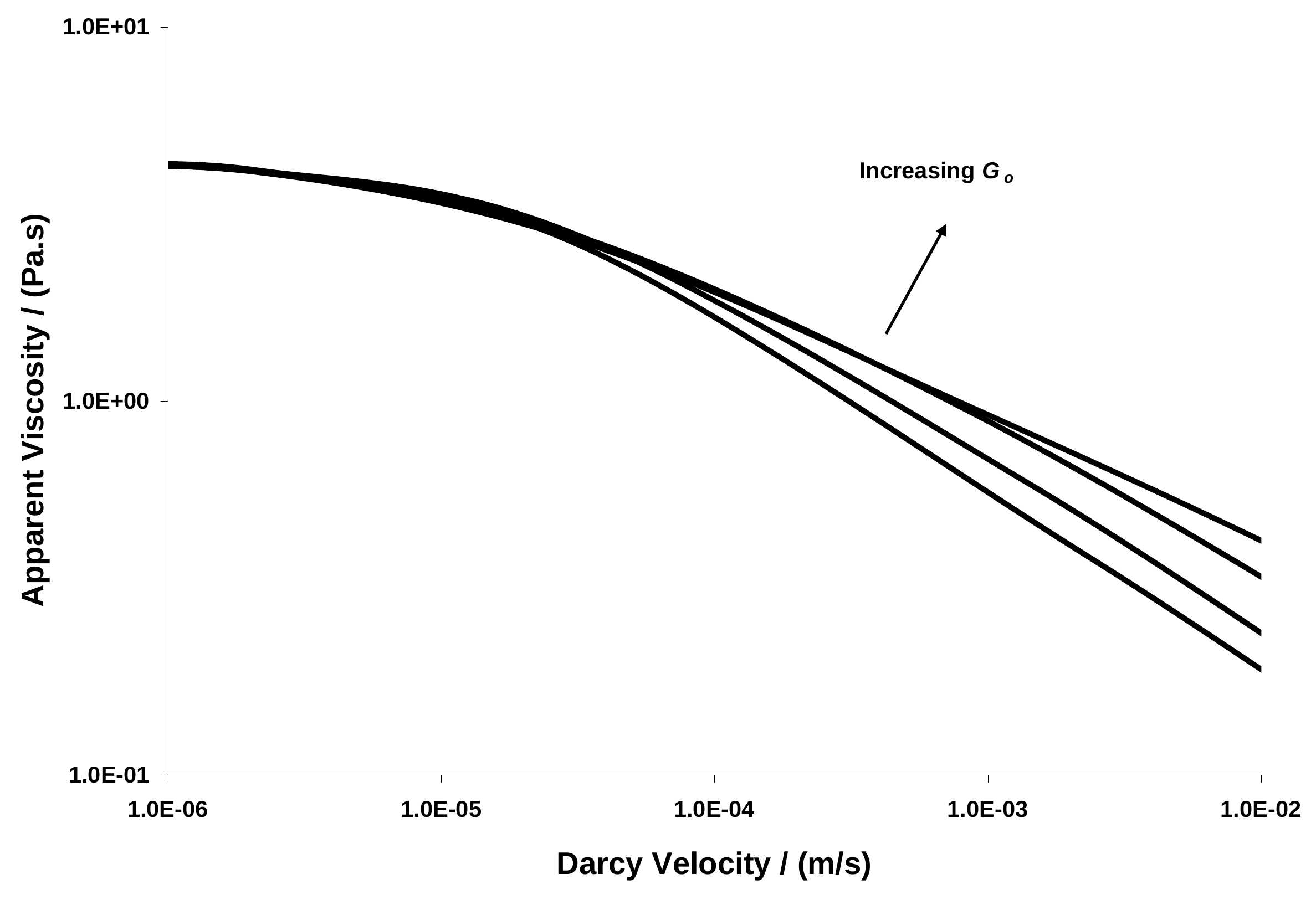}
  \caption[The \Tardy\ algorithm \sandp\ results for $\hVis$=0.001\,Pa.s, $\lVis$=1.0\,Pa.s, $\rxTimF$=1.0\,s,
            $\kF$=10$^{-5}$\,Pa$^{-1}$, $\fe$=1.0, $\fm$=0.5, $m$=10 slices, with varying $\Go$ (0.1, 1.0, 10 and 100\,Pa)]
  {The \Tardy\ algorithm \sandp\ results for $\hVis$=0.001\,Pa.s, $\lVis$=1.0\,Pa.s, $\rxTimF$=1.0\,s,
            $\kF$=10$^{-5}$\,Pa$^{-1}$, $\fe$=1.0, $\fm$=0.5, $m$=10 slices, with varying $\Go$ (0.1, 1.0, 10 and 100\,Pa).}
  \label{TardySP5}
\end{figure}
\begin{figure}[!h]
  \centering{}
  \includegraphics
  [scale=0.5]
  {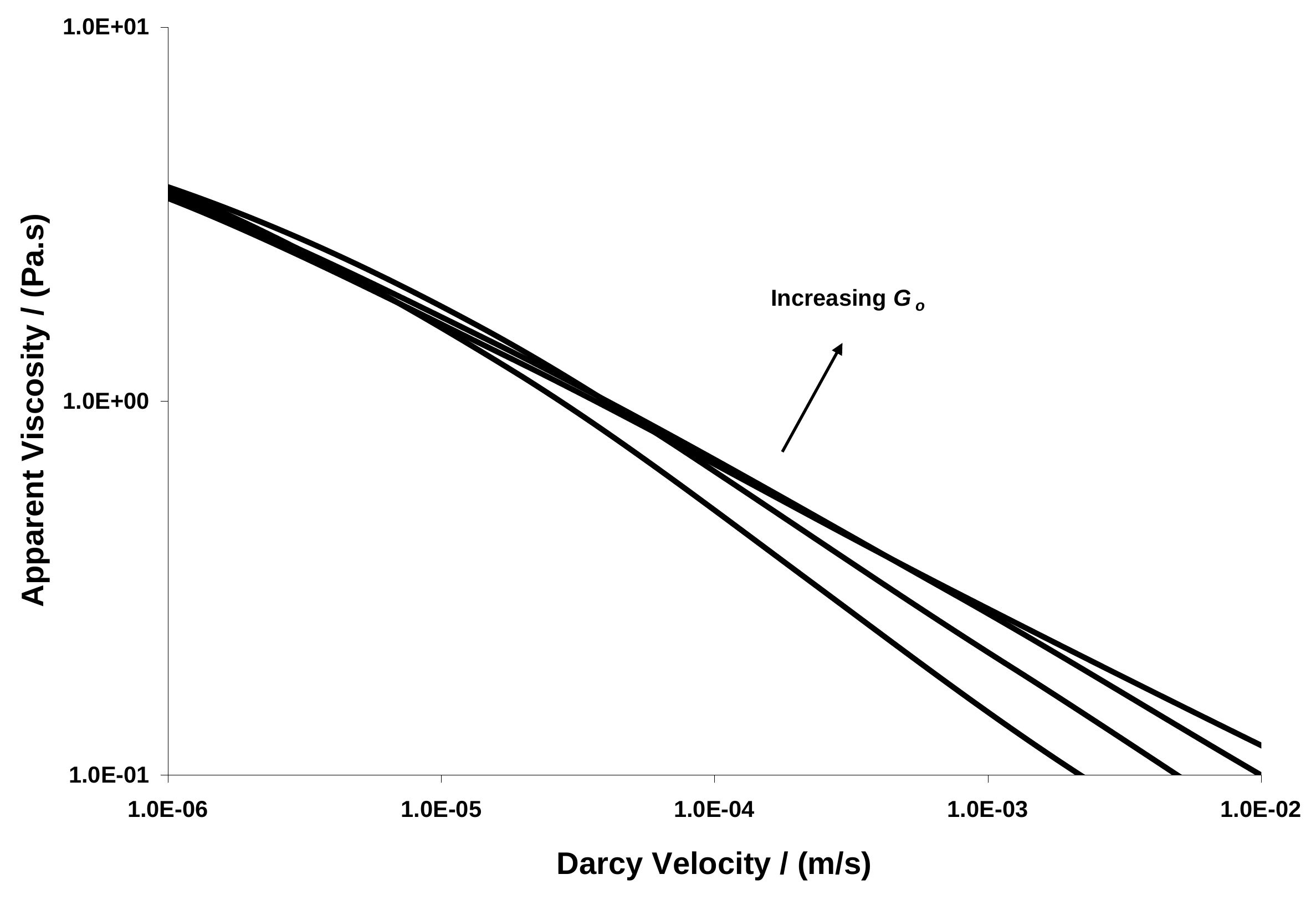}
  \caption[The \Tardy\ algorithm \Berea\ results for $\hVis$=0.001\,Pa.s, $\lVis$=1.0\,Pa.s, $\rxTimF$=1.0\,s,
            $\kF$=10$^{-5}$\,Pa$^{-1}$, $\fe$=1.0, $\fm$=0.5, $m$=10 slices, with varying $\Go$ (0.1, 1.0, 10 and 100\,Pa)]
  {The \Tardy\ algorithm \Berea\ results for $\hVis$=0.001\,Pa.s, $\lVis$=1.0\,Pa.s, $\rxTimF$=1.0\,s,
            $\kF$=10$^{-5}$\,Pa$^{-1}$, $\fe$=1.0, $\fm$=0.5, $m$=10 slices, with varying $\Go$ (0.1, 1.0, 10 and 100\,Pa).}
  \label{TardyB5}
\end{figure}

\subsubsection{Shear-Thickening}\label{}
A slight \shThik\ effect was observed when setting $\lVis < \hVis$ while holding the other
parameters constant. The effect of increasing $\Go$ on the apparent viscosities at high-shear rates
was similar to the effect observed for the \shThin\ fluids though it was at a smaller scale.
However, the low-shear viscosities are not affected, as in the case of \shThin\ fluids. A
convergence for the apparent viscosities at high-shear rates for large values of $\Go$ was also
observed as for \shThin. A sample of the results obtained in this investigation is presented in
Figures (\ref{TardySP6}) and (\ref{TardyB6}) for the \sandp\ and \Berea\ networks respectively on a
log-log scale. It should be remarked that as the \BauMan\ is originally a \shThin\ model, it should
not be expected to fully-predict \shThik\ phenomenon. However, the observed behavior is a good sign
for our model because as we push the algorithm beyond its limit, the results are still
qualitatively reasonable.

\begin{figure}[!t]
  \centering{}
  \includegraphics
  [scale=0.5]
  {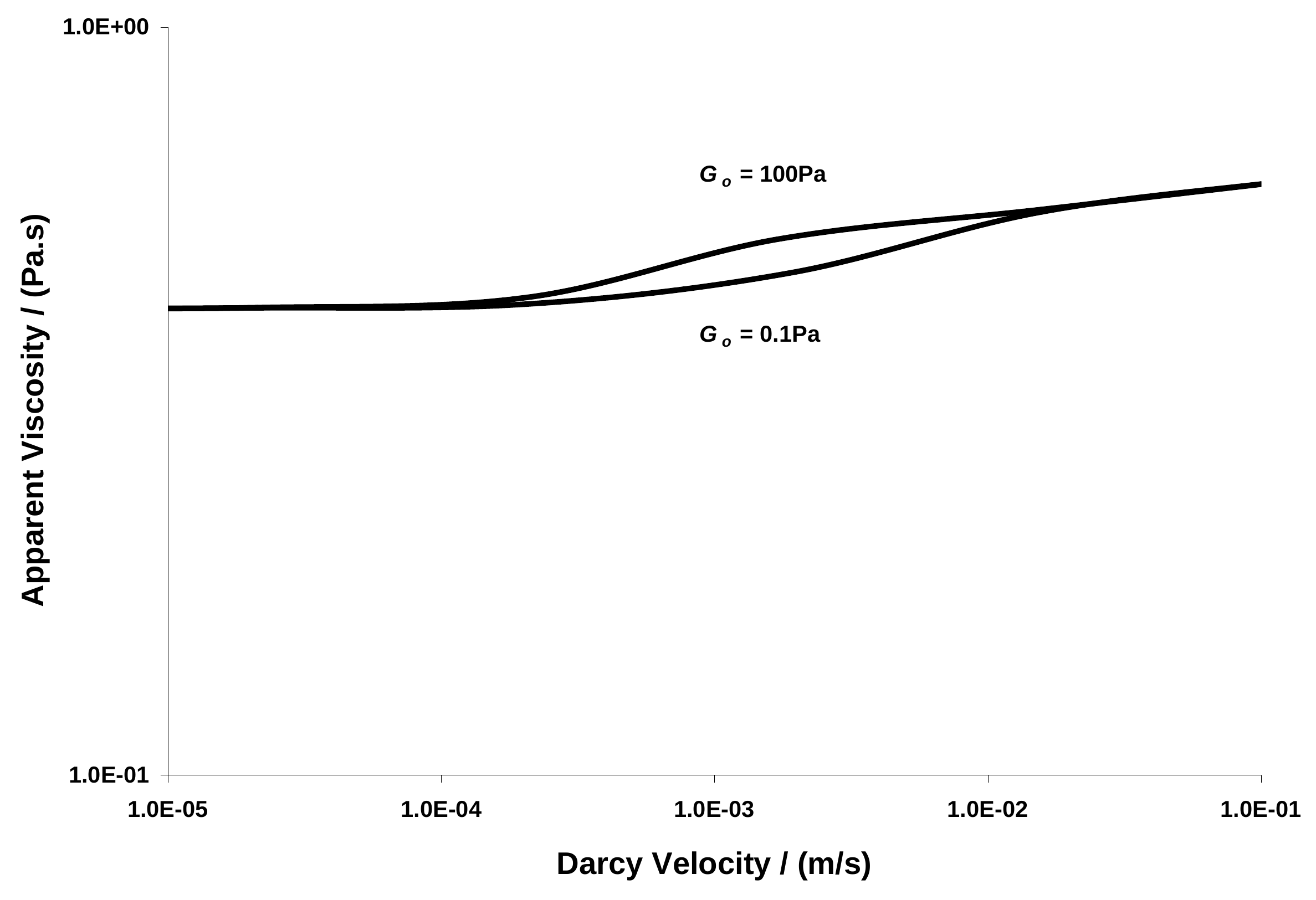}
  \caption[The \Tardy\ algorithm \sandp\ results for $\hVis$=10.0\,Pa.s, $\lVis$=0.1\,Pa.s, $\rxTimF$=1.0\,s,
            $\kF$=10$^{-4}$\,Pa$^{-1}$, $\fe$=1.0, $\fm$=0.5, $m$=10 slices, with varying $\Go$ (0.1 and 100\,Pa)]
  {The \Tardy\ algorithm \sandp\ results for $\hVis$=10.0\,Pa.s, $\lVis$=0.1\,Pa.s, $\rxTimF$=1.0\,s,
            $\kF$=10$^{-4}$\,Pa$^{-1}$, $\fe$=1.0, $\fm$=0.5, $m$=10 slices, with varying $\Go$ (0.1 and 100\,Pa).}
  \label{TardySP6}
\end{figure}
\begin{figure}[!h]
  \centering{}
  \includegraphics
  [scale=0.5]
  {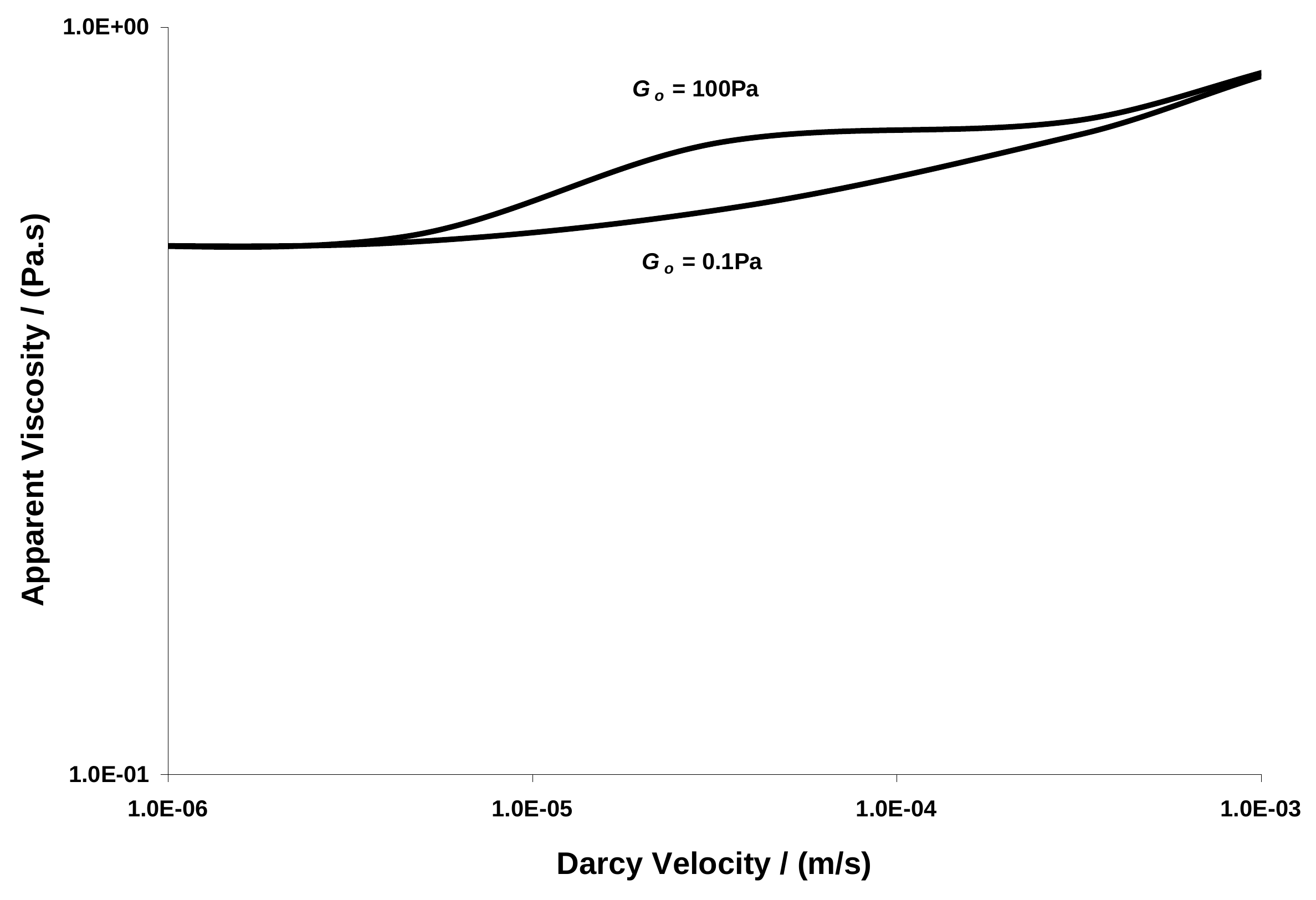}
  \caption[The \Tardy\ algorithm \Berea\ results for $\hVis$=10.0\,Pa.s, $\lVis$=0.1\,Pa.s, $\rxTimF$=1.0\,s,
            $\kF$=10$^{-4}$\,Pa$^{-1}$, $\fe$=1.0, $\fm$=0.5, $m$=10 slices, with varying $\Go$ (0.1 and 100\,Pa)]
  {The \Tardy\ algorithm \Berea\ results for $\hVis$=10.0\,Pa.s, $\lVis$=0.1\,Pa.s, $\rxTimF$=1.0\,s,
            $\kF$=10$^{-4}$\,Pa$^{-1}$, $\fe$=1.0, $\fm$=0.5, $m$=10 slices, with varying $\Go$ (0.1 and 100\,Pa).}
  \label{TardyB6}
\end{figure}

\subsubsection{Relaxation Time}\label{}
The effect of the structural relaxation time $\rxTimF$ was investigated for \shThin\ fluids by
varying this parameter over several orders of magnitude while holding the others constant. It was
observed that by increasing the structural relaxation time, the apparent viscosities were steadily
decreased. However, the decrease at high-shear rates has almost reached a saturation point where
beyond some limit the apparent viscosities converged to certain values despite a large increase in
$\rxTimF$. A sample of the results is given in Figure (\ref{TardySP7}) for the \sandp\ and Figure
(\ref{TardyB7}) for \Berea\ on a log-log scale. The effects of relaxation time are expected to ease
as the relaxation time increases beyond a limit such that the effect of interaction with capillary
constriction is negligible. Despite the fact that this feature requires extensive investigation for
quantitative confirmation, the observed behavior seems qualitatively reasonable considering the
flow regimes and the size of relaxation times of the sample results.

\begin{figure}[!t]
  \centering{}
  \includegraphics
  [scale=0.5]
  {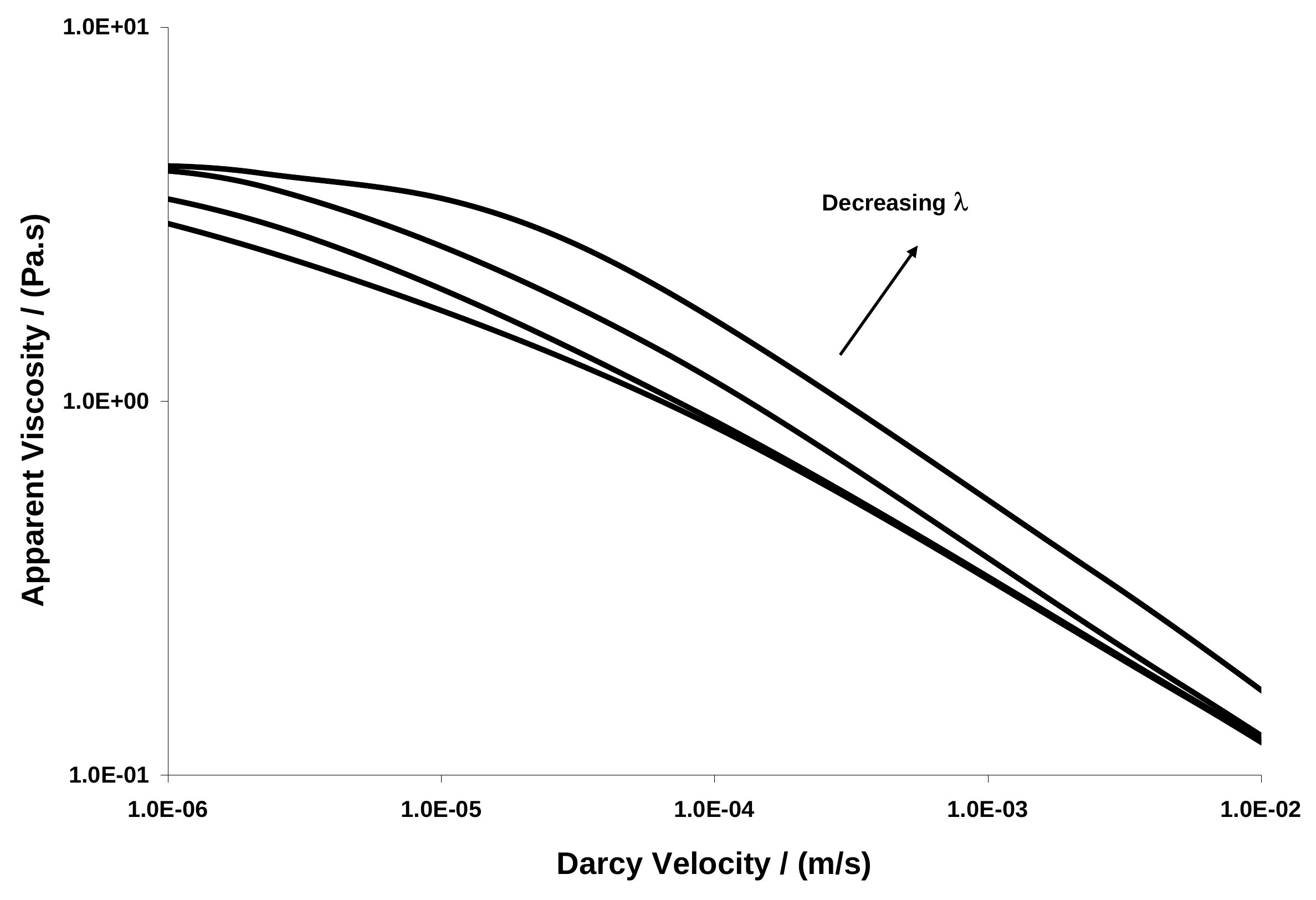}
  \caption[The \Tardy\ algorithm \sandp\ results for $\Go$=1.0\,Pa,  $\hVis$=0.001\,Pa.s, $\lVis$=1.0\,Pa.s,
            $\kF$=10$^{-5}$\,Pa$^{-1}$, $\fe$=1.0, $\fm$=0.5, $m$=10 slices, with varying $\rxTimF$ (0.1, 1.0, 10, and 100\,s)]
  {The \Tardy\ algorithm \sandp\ results for $\Go$=1.0\,Pa,  $\hVis$=0.001\,Pa.s, $\lVis$=1.0\,Pa.s,
            $\kF$=10$^{-5}$\,Pa$^{-1}$, $\fe$=1.0, $\fm$=0.5, $m$=10 slices, with varying $\rxTimF$ (0.1, 1.0, 10, and 100\,s).}
  \label{TardySP7}
\end{figure}
\begin{figure}[!h]
  \centering{}
  \includegraphics
  [scale=0.5]
  {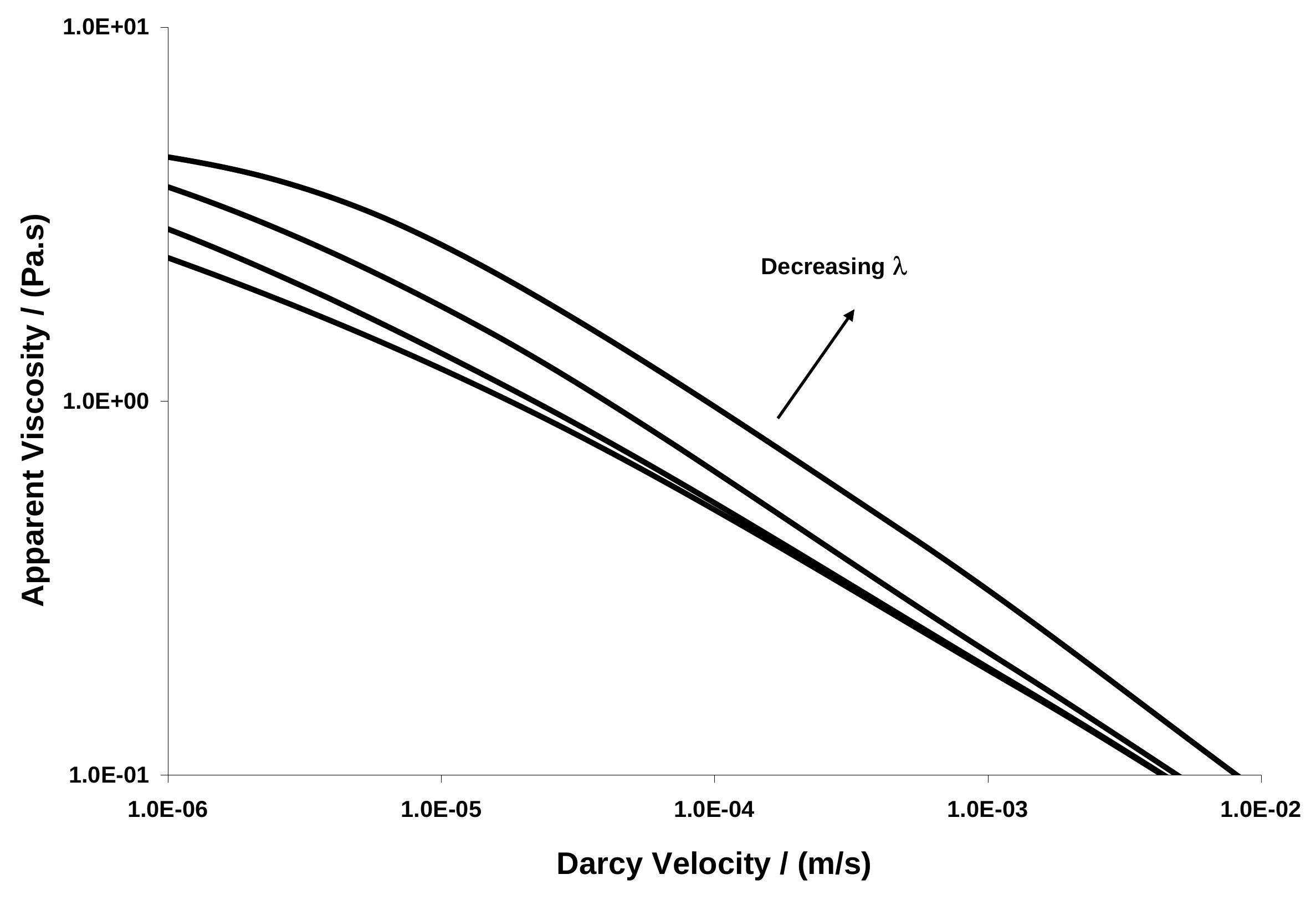}
  \caption[The \Tardy\ algorithm \Berea\ results for $\Go$=1.0\,Pa,  $\hVis$=0.001\,Pa.s, $\lVis$=1.0\,Pa.s,
            $\kF$=10$^{-5}$\,Pa$^{-1}$, $\fe$=1.0, $\fm$=0.5, $m$=10 slices, with varying $\rxTimF$ (0.1, 1.0, 10, and 100\,s)]
  {The \Tardy\ algorithm \Berea\ results for $\Go$=1.0\,Pa,  $\hVis$=0.001\,Pa.s, $\lVis$=1.0\,Pa.s,
            $\kF$=10$^{-5}$\,Pa$^{-1}$, $\fe$=1.0, $\fm$=0.5, $m$=10 slices, with varying $\rxTimF$ (0.1, 1.0, 10, and 100\,s).}
  \label{TardyB7}
\end{figure}

\subsubsection{Kinetic Parameter}\label{}
The effect of the kinetic parameter for structure break down $\kF$ was investigated for \shThin\
fluids by varying this parameter over several orders of magnitude while holding the others
constant. It was observed that by increasing the kinetic parameter, the apparent viscosities
generally decreased. However, in some cases the low-shear viscosities has not been substantially
affected. A sample of the results is given in Figures (\ref{TardySP8}) and (\ref{TardyB8}) for the
\sandp\ and \Berea\ networks respectively on a log-log scale. The decrease in apparent viscosity on
increasing the kinetic parameter is a natural response as the parameter quantifies structure break
down and hence reflects thinning mechanisms. It is a general trend that the low flow rate regimes
near the lower \NEW\ plateau are usually less influenced by \nNEW\ effects. It will therefore be
natural that the low-shear viscosities in some cases experienced minor changes.

\begin{figure}[!t]
  \centering{}
  \includegraphics
  [scale=0.5]
  {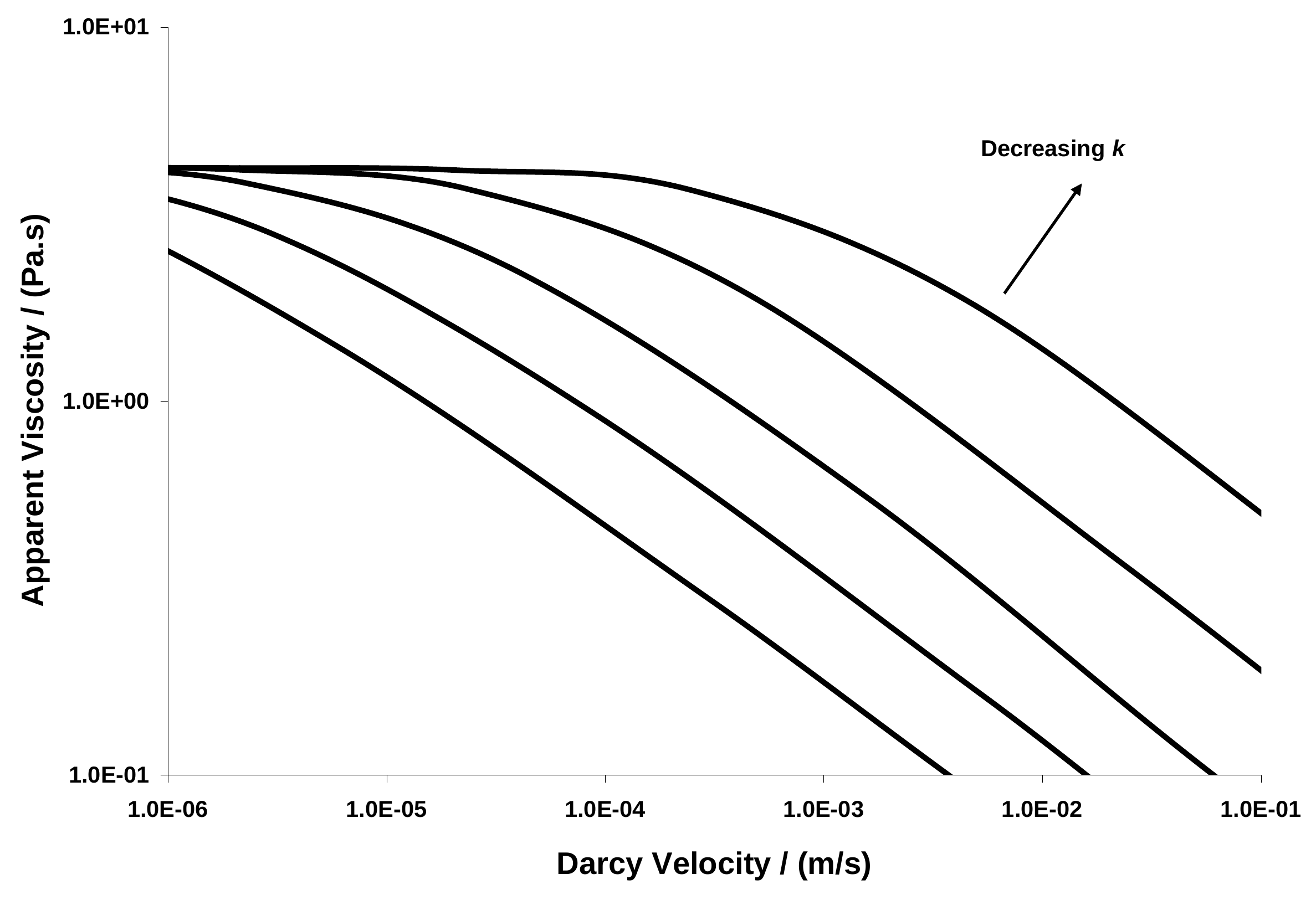}
  \caption[The \Tardy\ algorithm \sandp\ results for $\Go$=1.0\,Pa,  $\hVis$=0.001\,Pa.s, $\lVis$=1.0\,Pa.s,
            $\rxTimF$=10\,s, $\fe$=1.0, $\fm$=0.5, $m$=10 slices, with varying $\kF$ ($10^{-3}$,
            $10^{-4}$, $10^{-5}$, $10^{-6}$ and $10^{-7}$\,Pa$^{-1}$)]
  {The \Tardy\ algorithm \sandp\ results for $\Go$=1.0\,Pa,  $\hVis$=0.001\,Pa.s, $\lVis$=1.0\,Pa.s,
            $\rxTimF$=10\,s, $\fe$=1.0, $\fm$=0.5, $m$=10 slices, with varying $\kF$ ($10^{-3}$,
            $10^{-4}$, $10^{-5}$, $10^{-6}$ and $10^{-7}$\,Pa$^{-1}$).}
  \label{TardySP8}
\end{figure}
\begin{figure}[!h]
  \centering{}
  \includegraphics
  [scale=0.5]
  {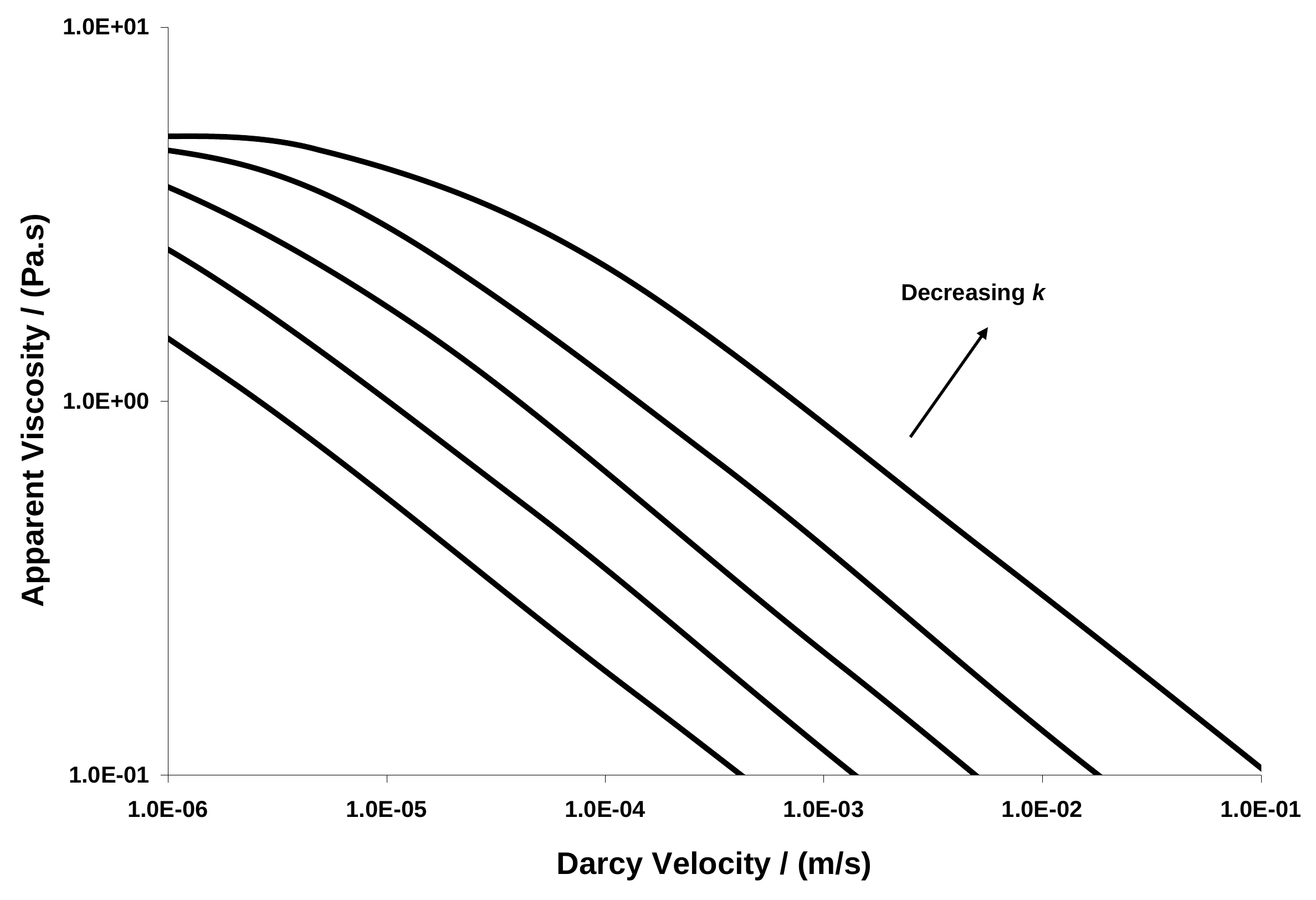}
  \caption[The \Tardy\ algorithm \Berea\ results for $\Go$=1.0\,Pa,  $\hVis$=0.001\,Pa.s, $\lVis$=1.0\,Pa.s,
            $\rxTimF$=10\,s, $\fe$=1.0, $\fm$=0.5, $m$=10 slices, with varying $\kF$ ($10^{-3}$,
            $10^{-4}$, $10^{-5}$, $10^{-6}$ and $10^{-7}$\,Pa$^{-1}$)]
  {The \Tardy\ algorithm \Berea\ results for $\Go$=1.0\,Pa,  $\hVis$=0.001\,Pa.s, $\lVis$=1.0\,Pa.s,
            $\rxTimF$=1.0\,s, $\fe$=1.0, $\fm$=0.5, $m$=10 slices, with varying $\kF$ ($10^{-3}$,
            $10^{-4}$, $10^{-5}$, $10^{-6}$ and $10^{-7}$\,Pa$^{-1}$).}
  \label{TardyB8}
\end{figure}

\clearpage

\section{Conclusions} \label{Conclusions}

A \nNEW\ flow simulation code based on pore-scale network modeling has been extended to account for
\vc\ and \thixotropic\ behavior using a \BauMan\ fluid model. The basis of the implementation is a
numerical algorithm originally suggested by Philippe \Tardy. A modified version of this algorithm
was used to investigate the effect of \convdiv\ geometry on the \steadys\ flow. The implementation
of this model has been examined and assessed using a \sandp\ and a \Berea\ networks. The generic
behavior indicates qualitatively correct trends. A conclusion was reached that the current
implementation is a proper basis for investigating the \steadys\ \vc\ effects and some
\thixotropic\ effects due to \convdiv\ geometries in porous media. This implementation can be used
as a suitable foundation for further development. The future work can include elaborating the
modified \Tardy\ algorithm and thoroughly investigating its \vc\ and \thixotropic\ predictions in
quantitative terms. The code (called Non-Newtonian Code 2) with complete documentation and the
networks can be obtained from this URL:
\url{http://www3.imperial.ac.uk/earthscienceandengineerinresearch/perm/porescalemodellinsoftware/non-newtonian\%20code}.

} 


\newpage
\phantomsection \addcontentsline{toc}{section}{Acknowledgements} \noindent
{\LARGE \bf \vspace{3.0cm} \\ Acknowledgements} \vspace{0.5cm}\\
I would like to thank Dr Valerie Anderson and Dr John Crawshaw of Schlumberger Cambridge Research
Center, and Prof. Martin Blunt and Prof. Geoffrey Maitland of Imperial College London for their
help and advice with regards to various aspects related to the materials in this article.


\newpage

\phantomsection \addcontentsline{toc}{section}{Nomenclature}
\noindent \vspace{1.0cm} \\
{\LARGE \bf Nomenclature}
\vspace{0.7cm} 

\begin{supertabular}{ll}
$\sR$                 & strain rate (s$^{-1}$) \\
$\rsTen$              & rate-of-strain tensor \\
$\rxTimF$             & structural relaxation time in \FRED\ model (s) \\
$\rxTim$              & relaxation time (s) \\
$\rdTim$              & retardation time (s) \\
$\Vis$                & viscosity (Pa.s) \\
$\lVis$               & zero-shear viscosity (Pa.s) \\
$\hVis$               & infinite-shear viscosity (Pa.s) \\
$\sS$                 & stress (Pa) \\
$\sTen$               & stress tensor \\

\\

$\fe$                 & scale factor for the entry of corrugated tube (---) \\
$\fm$                 & scale factor for the middle of corrugated tube (---) \\
$G$                   & geometric conductance (m$^4$) \\
$G'$                  & flow conductance (m$^3$.Pa$^{-1}$.s$^{-1}$)\\
$\Go$                 & elastic modulus (Pa) \\
$\kF$                 & parameter in \FRED\ model (Pa$^{-1}$) \\
$L$                   & tube length (m) \\
$P$                   & pressure (Pa) \\
$\Delta P$            & pressure drop (Pa) \\
$Q$                   & volumetric flow rate (m$^{3}$.s$^{-1}$) \\
$r$                   & radius (m) \\
$R$                   & tube radius (m) \\
$R_{eq}$              & equivalent radius (m) \\
$\Rmax$               & maximum radius of corrugated capillary \\
$\Rmin$               & minimum radius of corrugated capillary \\
$t$                   & time (s) \\
$\fVel$               & fluid velocity vector \\
$\nabla \fVel$        & fluid velocity gradient tensor \\
$V$                   & fluid speed (m.s$^{-1}$) \\
$\delta x$            & small change in $x$ (m) \\

\\

{$\ucd \cdot$}          & upper convected time derivative \\
$(\cdot)^{T}$           & matrix transpose \\
$x_{_{l}}$             & network lower boundary in the \nNEW\ code \\
$x_{_{u}}$             & network upper boundary in the \nNEW\ code \\
$\verb|       |$        &  \\
\end{supertabular}

\vspace{0.5cm}

\noindent %
{\bf Note}: units, when relevant, are given in the SI system. Vectors and tensors are marked with
boldface. Some symbols may rely on the context for unambiguous identification.


\newpage

\phantomsection \addcontentsline{toc}{section}{References} %
\bibliographystyle{unsrt}    
\bibliography{Biblio}        



\newpage

\phantomsection \addcontentsline{toc}{section}{Appendix A: \ConvDiv\ Geometry} %
\begin{spacing}{2.5}
{\LARGE \bf \noindent Appendix A: \ConvDiv\ Geometry and Tube Discretization}
\end{spacing}

\vspace{1.0cm}

\begin{figure}[!t]
  \centering{}
  \includegraphics
  [scale=0.35]
  {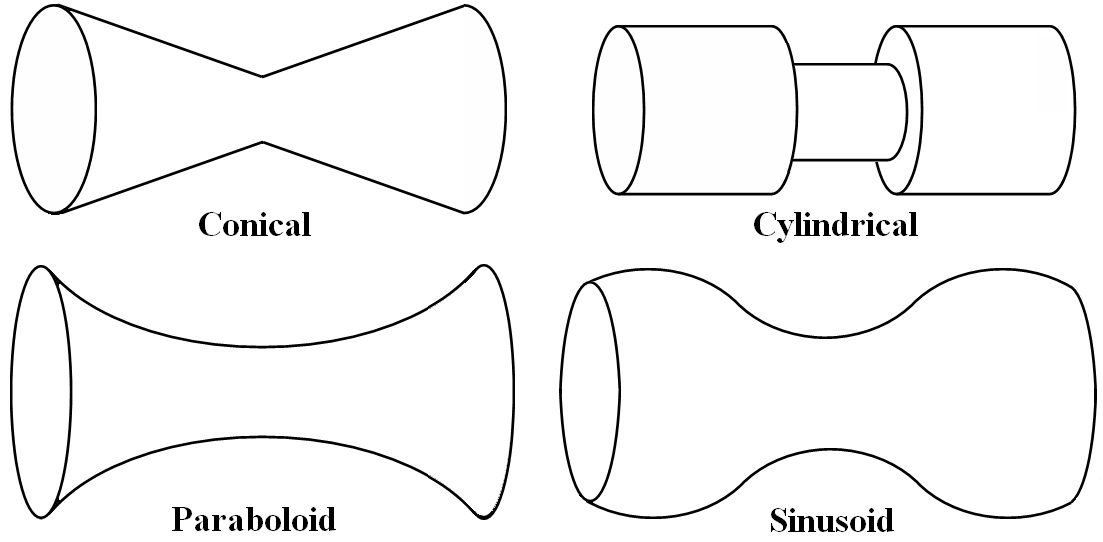}
  \caption[Examples of corrugated capillaries which can be used to model \convdiv\ geometry in porous media]
  {Examples of corrugated capillaries which can be used to model \convdiv\ geometry in porous media.}
  \label{ConvDivGeom}
\end{figure}

\begin{figure}[!t]
  \centering{}
  \includegraphics
  [scale=0.55]
  {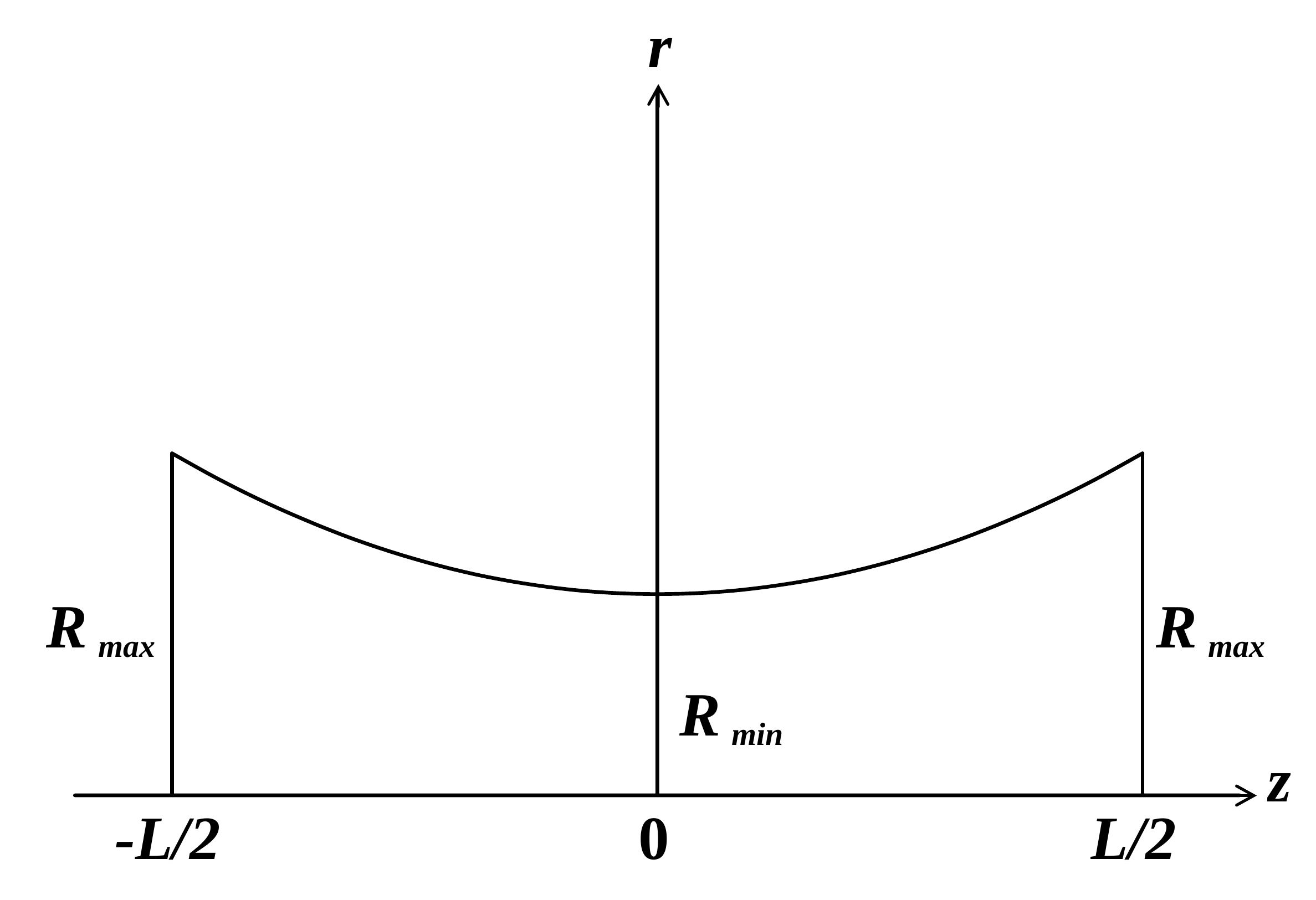}
  \caption[Radius variation in the axial direction for a corrugated paraboloid capillary using a cylindrical coordinate system]
  {Radius variation in the axial direction for a corrugated paraboloid capillary using a cylindrical coordinate system.}
  \label{ParaboloidTube}
\end{figure}

\noindent There are many simplified \convdiv\ geometries for corrugated capillaries which can be
used to model the flow of \vc\ fluids in porous media; some suggestions are presented in Figure
(\ref{ConvDivGeom}). In this study we adopted the paraboloid and hence developed simple formulae to
find the radius as a function of the distance along the tube axis, assuming cylindrical coordinate
system, as shown in Figure (\ref{ParaboloidTube}).

For the paraboloid depicted in Figure (\ref{ParaboloidTube}), we have
\begin{equation}\label{paraboloidRad}
    r(z) = a z^{2} + b z + c
\end{equation}

Since
\begin{equation}\label{paraboloidPoints}
    r(-L/2) = r(L/2) = R_{max}     \verb|       |     \&     \verb|       |     r(0) = R_{min}
\end{equation}
the paraboloid is uniquely identified by these three points. On substituting and solving these
equations simultaneously, we obtain
\begin{equation}\label{paraboloidCoef}
    a = \left(\frac{2}{L}\right)^{2} (R_{max} - R_{min})  \verb|       |  b = 0
    \verb|       |  \&     \verb|       |  c = R_{min}
\end{equation}
that is
\begin{equation}\label{paraboloidFinal}
    r(z) = \left(\frac{2}{L}\right)^{2} (R_{max} - R_{min}) z^{2} + R_{min}
\end{equation}

In the modified \Tardy\ algorithm for \steadys\ \vc\ flow as implemented in our \nNEW\ code, when a
capillary is discretized into $m$ slices the radius $r(z)$ is sampled at $m$ $z$-points given by
\begin{equation}\label{radiusPoints}
    z = -\frac{L}{2} + k \frac{L}{m}
    \verb|       | (k = 1,...,m)
\end{equation}

More complex polynomial and sinusoidal \convdiv\ profiles can also be modeled using this approach.


\end{document}